\documentclass{article}
\usepackage{array}
\usepackage{algorithm}
\usepackage{algorithmicx}
\usepackage{algpseudocode}
\usepackage{float}
\usepackage{gensymb}
\usepackage{graphicx}
\usepackage[utf8]{inputenc}
\usepackage[backend=biber, style=nature, maxbibnames=5]{biblatex}
\usepackage{amsfonts}
\usepackage{enumerate}
\usepackage{amsmath}
\usepackage{comment}
\usepackage{appendix}
\usepackage{booktabs}
\usepackage[labelfont=bf]{caption} % makes Figure in caption bold
\usepackage{changepage}
\usepackage{color}
\usepackage{authblk}
\usepackage{amssymb}
\usepackage{lineno}
\usepackage{longtable}
\usepackage{multirow}
\usepackage{physics}
\usepackage[onehalfspacing]{setspace}
\usepackage{stackengine} 
\usepackage{wrapfig}
\usepackage{url}
\usepackage[table,xcdraw]{xcolor}
\usepackage{tikz}
\usepackage{appendix} % [page,header]
\usepackage{titletoc}

\usepackage[colorlinks=true,citecolor=blue,urlcolor=blue]{hyperref}		\hypersetup{linkcolor=blue}

\usepackage[a4paper,margin=1.5cm]{geometry}
\newcommand{\of}[1]{\left(#1\right)}

\newcommand{\uf}[1]{\left\{#1\right\}}

\DeclareMathOperator{\unif}{Unif}

\newcommand \stacksymbol[3]{\mathrel{\stackunder[2pt]{\stackon[4pt]{$#1$}{$\scriptscriptstyle#2$}}{$\scriptscriptstyle#3$}}}

\DeclareRobustCommand{\circled}[2][1pt]{\tikz[baseline=(char.base)]{
            \node[shape=circle,draw,inner sep= #1] (char) {#2};}}

\definecolor{darkgreen}{rgb}{0.06, 0.6, 0.3}
\definecolor{dblue}{RGB}{6,69,173}
\definecolor{burgundy}{rgb}{0.5, 0.0, 0.13}
\definecolor{mulberry}{rgb}{0.7831, 0.0, 0.6549}

\usepackage{fancyhdr}
\fancypagestyle{plain}{%
    \fancyhf{}%
    \fancyfoot[R]{\thepage}%
}

\usepackage{filecontents}

\begin{document}

\title{ \vspace*{0cm} \textbf{Stochastic modelling reveals that chromatin folding buffers epigenetic landscapes against sirtuin depletion during DNA damage}}

\author[1,*]{Daria Stepanova}%
\author[2,3]{Helen M. Byrne}%
\author[4,1,5]{Tom\'{a}s Alarc\'{o}n}%

\affil[1]{Centre de Recerca Matem\`{a}tica, Bellaterra (Barcelona), Spain}%
\affil[2]{Wolfson Centre for Mathematical Biology, Mathematical Institute, University of Oxford, Oxford, UK}%
\affil[3]{Ludwig Institute for Cancer Research, Nuffield Department of Medicine, University of Oxford, Oxford, UK}
\affil[4]{Instituci\'{o} Catalana de Recerca i Estudis Avan\c{c}ats, Barcelona, Spain}%
\affil[5]{Departament de Matem\`{a}tiques, Universitat Aut\`{o}noma de Barcelona, Bellaterra (Barcelona), Spain}%

% \affil[$\dagger$]{These authors contributed equally to this work.}
\affil[*]{\url{dstepanova@crm.cat}}

\date{\today}
\begingroup
\let\center\flushleft
\let\endcenter\endflushleft
\maketitle
\endgroup

% 150 words
\begin{abstract}
Epigenetic landscapes, represented by patterns of chemical modifications on histone tails, are essential for maintaining cell identity and tissue homeostasis. These landscapes are shaped by multiple factors, including local biochemical signals and the three-dimensional organisation of chromatin. However, their response to genomic stress, such as DNA double-strand breaks (DSBs), remains incompletely understood. Here, we use a stochastic model of histone modification dynamics integrated with chromatin architecture to investigate how local depletion of sirtuins, histone deacetylases involved in DSB repair, destabilises epigenetic patterns. Our simulations recapitulate experimental findings in which sirtuin relocalisation to DSB sites leads to the epigenetic erosion and suggest that the resulting landscape depends on enzyme levels and chromatin geometry. Importantly, chromatin regions with large domains of long-range contacts are more resilient to epigenetic destabilisation. These findings suggest that chromatin folding can buffer against relocation of histone-modifying enzymes, highlighting a structural mechanism for preserving epigenetic integrity under stress.
\end{abstract}

\section{Introduction}

Epigenetics refers to heritable but reversible modifications that regulate gene expression without altering the underlying DNA sequence \supercite{felsenfeld2014brief}. Among the most extensively studied epigenetic modifications are post-translational modifications of histone proteins, particularly those on the terminal sequences (tails) of histone H3, which influence chromatin accessibility and, thereby, control transcriptional activity \supercite{gross2015chromatin}. These modifications, which include methylation and acetylation, are deposited, removed, and read by specialised enzymes, known collectively as histone-modifying enzymes \supercite{marmorstein2009histone}. In this way, epigenetic marks play a crucial role in establishing and maintaining cellular identity, ensuring that the correct genes are active or repressed in a particular cell type.

Epigenetic regulation occurs in the context of hierarchical folding of the chromatin fibre, which compacts the extensive DNA sequence to fit within the cell nucleus and also provides the structural and functional framework necessary for establishing stable epigenetic signatures and, consequently, cell identities. At the most fundamental level, DNA is wrapped around nucleosomes, octameric complexes composed of four core histone types, forming the chromatin polymer \supercite{gross2015chromatin}. This polymer is further organised into higher-order architectures by structural proteins such as CTCF and the cohesin complex \supercite{grubert2020landscape}. These proteins mediate the formation of chromatin loops and topologically associating domains (TADs), which range in size from kilobases (kb) to several megabases (Mb) \supercite{grubert2020landscape}. Such higher-order domains facilitate spatial proximity between regulatory elements, such as promoters and enhancers, thereby enabling gene activation, while also insulating domains from one another to restrict aberrant interactions and maintain proper genomic regulation.

The relationship between chromatin conformation and epigenetic regulation is fundamentally bidirectional. On the one hand, specific histone modifications directly influence chromatin compaction and its folding patterns \supercite{gross2015chromatin}. Trimethylation of lysine 27 on histone H3 (H3K27me3) is a repressive modification associated with tightly packed, transcriptionally silent chromatin, whereas acetylation of the same residue (H3K27ac) is an activating mark linked to open chromatin and active gene transcription. Conversely, the three-dimensional spatial organisation of the genome, defined by its physical folding and long-range contacts, shapes how histone-modifying enzymes access and modify chromatin. These enzymes often contain reader domains that can recognise pre-existing histone marks, which, in turn, allosterically enhance the activity of their writer domains to deposit new modifications at nearby loci. Such feedback loops can reinforce and stabilise local epigenetic states \supercite{sneppen2019theoretical}. Together, these coupled processes generate what is referred to as epigenetic landscapes -- complex, heterogeneous patterns of activating and repressive marks across the genome that encode and maintain cell identity.

The stability of these epigenetic landscapes is crucial. Disruptions in their organisation can lead to the loss of cell identity, altered gene expression, and the onset of diseases, including cancer and neurodegenerative disorders \supercite{zoghbi2016epigenetics,tollefsbol2018epigenetics,zhang2020epigenetics}. The concept of a heterogeneous epigenetic landscape, characterised by locally stable domains of chromatin states, has been particularly influential in developmental biology and pathology, as it offers a framework to understand how cells maintain their identity and how their identity might be reset or lost \supercite{perino2016chromatin,pujadas2012regulated}.

A recent example of the biological relevance of epigenetic landscape stability is provided by the work of Yang et al.\supercite{yang2023loss}, who challenged the long-standing assumption that ageing is primarily driven by the accumulation of genetic mutations. In their study, the authors presented evidence that ageing might instead result from the loss of epigenetic information, particularly in response to DNA damage. Using a mouse model and a system known as ICE (inducible changes to the epigenome), they showed that a severe form of DNA damage, such as double-strand breaks (DSBs), could lead to persistent alterations in the epigenetic profile of cells. Importantly, this erosion of the epigenetic landscape occurred without significant changes in the underlying DNA sequence since induced DSBs were non-mutagenic \supercite{yang2023loss}.

Central to their hypothesis are sirtuins, a family of NAD (nicotinamide adenine dinucleotide)-dependent histone deacetylases involved in both chromatin regulation and DNA repair \supercite{yang2023loss}. Upon DNA damage, sirtuins are recruited to DSB sites to facilitate repair processes. However, their relocation comes at a cost: their absence from chromatin allows activating marks such as H3K27ac to accumulate, potentially destabilising repressive domains. If the original epigenetic patterns are not restored post-repair, this can lead to long-term changes in chromatin organisation and gene expression, contributing to cellular ageing \supercite{yang2023loss}. Yang et al. also demonstrated that resetting the epigenetic landscape using a backup youthful profile can partially reverse these ageing phenotypes, emphasising the therapeutic potential of targeting epigenetic stability.

While these findings suggest a general mechanism by which DNA damage can destabilise epigenetic landscapes, certain aspects of this process remain poorly understood. In particular, it is unclear whether all epigenetic landscapes deteriorate in the same way or whether their stability depends on specific features of the chromatin environment, including the local availability of regulatory enzymes and pre-existing histone modifications. Moreover, it remains to be determined whether all chromatin architectures are equally susceptible to erosion induced by elevated levels of DSBs or whether certain chromatin geometries can mitigate the effects of sirtuin sequestration and enhance the robustness of the underlying epigenetic landscape.

To investigate these questions, we extend a recently proposed mathematical model of epigenetic regulation and chromatin architecture \supercite{stepanova2025understanding}. This model provides a mesoscopic, stochastic framework for exploring the dynamic interplay between three-dimensional genome organisation and histone modification processes. It focuses on covalent modifications of histone H3 at lysines 4 and 27, specifically the activating marks H3K4me3 and H3K27ac, as well as the repressive mark H3K27me3. Central to the model are reader-writer feedback mechanisms governing the propagation of these marks \supercite{sneppen2019theoretical,stepanova2025understanding}. By incorporating chromatin architecture through a contact-frequency matrix that quantifies spatial interactions between genomic loci, the model enables simulation of how genome folding influences the establishment and stability of epigenetic states.

In this work, we extend the model by Stepanova et al. to incorporate the effect of DSB-induced enzyme sequestration, motivated by the ICE system and other damage-related cellular contexts. In particular, we simulate scenarios in which sirtuin availability is reduced due to relocation to DNA repair sites, and examine how this affects the stability of epigenetic landscapes. We investigate several biologically relevant conditions, including varying damage loads, local chromatin geometries, and baseline levels of histone-modifying enzymes.

Our results show that the relocation of enzymes to DNA repair sites can indeed destabilise pre-existing epigenetic domains. In regimes of low sirtuin availability, we observe the flattening of previously heterogeneous landscapes into uniformly open, hyperacetylated states. Conversely, under high sirtuin availability, initially silenced chromatin can become more permissive due to a partial loss of repressive marks, replaced by activating acetylating modifications. These outcomes highlight the role of enzyme availability as a key control parameter in shaping epigenetic states and the erosion mechanisms. Furthermore, our analysis and simulations reveal that the resilience of the epigenetic landscape depends on chromatin architecture. Larger or more dynamically interacting domains are more robust to disruption, even when composed of relatively weak contact intensities. In contrast, smaller or fragmented domains are more susceptible to epigenetic erosion under the same levels of DNA damage. %Notably, these findings suggest that the chromatin reorganisation observed in ICE-treated cells, such as the formation of new contacts between enhancers and promoters across TAD boundaries \supercite{yang2023loss} may represent an adaptive strategy to preserve or restore regulatory landscapes. By enlarging functional domains or increasing connectivity across them, cells may compensate for the destabilising effects of DNA damage and enzyme redistribution.

The rest of the paper is organised as follows. Section~\ref{main:results} begins with a summary of our modelling approach. In Section~\ref{main:results1}, we employ theoretical analysis and simulations to investigate how different levels of enzyme availability influence the stability and transformation of epigenetic patterns. In Section~\ref{main:results2}, we focus on the role of chromatin geometry and show how domain size, internal connectivity, and spanning interactions influence a landscape's sensitivity to destabilisation. In Section~\ref{main:results3}, we confirm these findings using stochastic simulations with more realistic chromatin conformations. We conclude in Section~\ref{main:discussion} with a summary of our results and a discussion of the broader implications of the work. Lastly, in Section~\ref{main:methods1}, we summarise our modelling framework, including the baseline mesoscopic model of chromatin-epigenetic interaction, its extension to include enzyme sequestration by DSBs, and the choice of parameter regimes based on experimentally observed DNA damage frequencies in both physiological and pathological conditions. 

\begin{figure}[!t]
\begin{adjustwidth}{-0.5cm}{-0.5cm}
\begin{center}
\includegraphics[width=19cm]{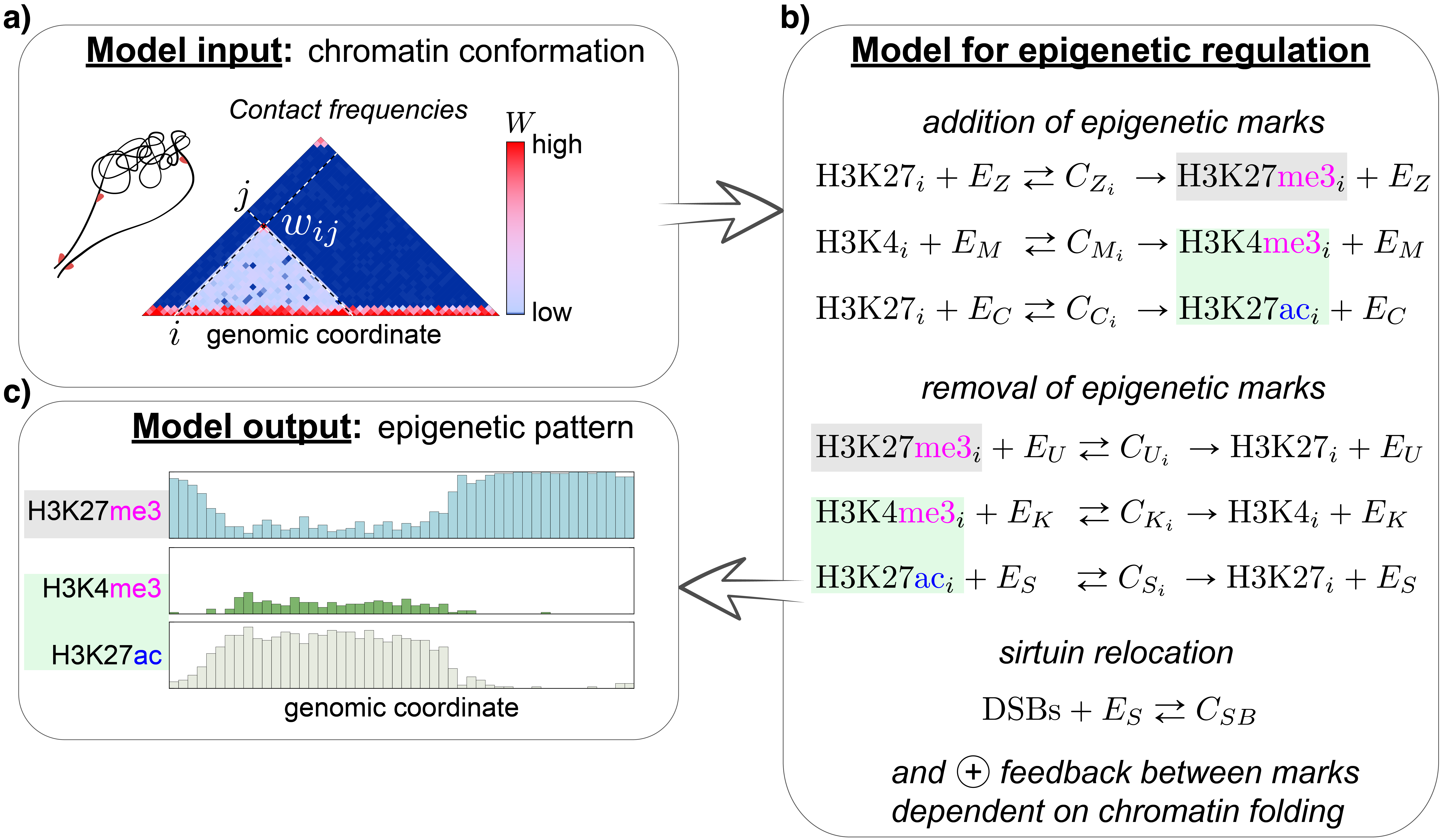}
\caption{\textbf{A schematic of our modelling approach.} \textbf{a)} Our model for epigenetic regulation incorporates chromatin geometry, quantified via a contact frequency map that describes pairwise interaction intensities between genomic loci. High values in this matrix, $W$, indicate frequent spatial proximity. Diagonal elements, $w_{ii}$, with high values represent local interactions along the chromatin fiber, while off-diagonal elements, $w_{ij}$, $i \neq j$, with high values highlight higher-order structures such as loops or domains. \textbf{b)} The chromatin contact matrix $W$ is integrated into our model of epigenetic mark dynamics. The model tracks the addition and removal of one repressive modification (H3K27me3, grey) and two activating modifications (H3K4me3 and H3K27ac, green), catalysed by distinct enzymes ($E_J$, where $J$ denotes enzyme type). Chromatin conformation influences these modifications by intensifying or inhibiting the feedback mechanisms between the existing marks and the formation of new modifications. This is implemented in our model by assuming that the rate constants of substrate-enzyme complex formation ($C_{J_i}$) are functions of the epigenetic marks at other genomic sites weighted by the frequency of interaction between them ($W$). Sirtuins ($E_S$), which participate in removing acetyl marks from histones, can also be recruited to repair damaged DNA. This is incorporated into the model by assuming that these enzymes can form a complex with a DSB repair site (represented by a dimensionless parameter $B$ in our model; see also Section~\ref{main:methods12}). A detailed model description is provided in the Supplementary Material of this work and in reference\supercite{stepanova2025understanding}. \textbf{c)} By employing a quasi-steady state approximation, the enzymatic reactions in \textbf{b} can be simplified to track the evolution of epigenetic mark distribution along the chromatin region. Due to feedback mechanisms, the two activating marks are typically positively correlated, while both are negatively correlated with the repressive modification H3K27me3.}
\label{Figure1}
\end{center}
\end{adjustwidth}
\end{figure}

\section{Results} \label{main:results}

We build upon a previously developed stochastic model of epigenetic regulation that integrates site-specific histone dynamics with the three-dimensional (3D) architecture of chromatin folding\supercite{stepanova2025understanding}. In this model, chromatin is represented as a linear fibre of genomic sites, each of which can acquire or lose specific epigenetic marks depending on the activity of local and spatially interacting enzymes. The chromatin conformation is encoded through a contact matrix derived from folding geometries of pairwise spatial interactions between genomic sites (Fig.~\ref{Figure1}a). The model tracks three key histone modifications, H3K27me3, H3K4me3, and H3K27ac, each associated with distinct regulatory outcomes, and whose addition and removal are mediated by six enzyme classes (Fig.~\ref{Figure1}b).

To reduce the model's complexity while preserving its key dynamics, we apply a quasi-steady-state approximation valid under low enzyme abundance. This allows us to coarse-grain the model and track the temporal evolution of the three epigenetic marks at each genomic coordinate (Fig.~\ref{Figure1}c). The resulting framework enables us to capture how chromatin structure modulates the emergence of heterogeneous epigenetic patterns, characterised by the coexistence of silenced and active domains (Fig.~\ref{Figure1}c).

To explore how DNA damage affects these patterns, we extend the model by incorporating the redistribution of sirtuins, histone deacetylases involved in both epigenetic regulation and DSB repair\supercite{yang2023loss}. Specifically, we include a reversible reaction to account for the sequestration of free sirtuins to DNA DSBs (see Fig.~\ref{Figure1}b), thereby reducing their availability for histone deacetylation. This extended model allows us to study how increasing levels of DSBs can destabilise existing epigenetic landscapes.

More detail on our model formulation can be found in Section~\ref{main:methods1} and Supplementary Material of this work.

\subsection{Erosion of epigenetic landscapes in high levels of double-strand breaks} \label{main:results1}

\begin{figure}[!t]
\begin{adjustwidth}{-0.5cm}{-0.5cm}
\begin{center}
\includegraphics[width=17cm]{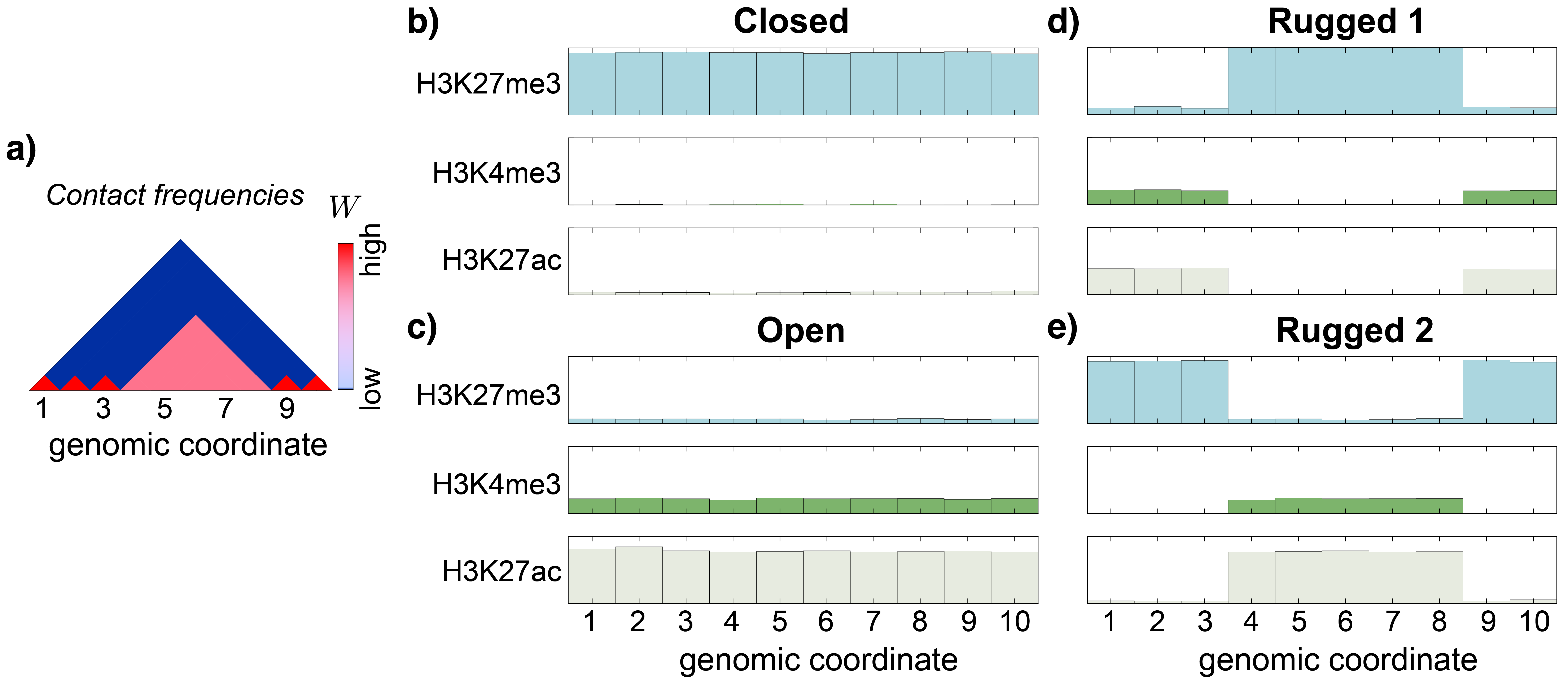}
\caption{\textbf{Contact matrix and emergent epigenetic patterns in a two-region chromatin model.} \textbf{a)} Schematic representation of the contact frequency matrix, $W$, for a two-region chromatin system. One region (sites 1-3 and 9-10) exhibits only local (diagonal) interactions of strength $w_{diag}$, while the central TAD-like region (sites 4-8) displays uniform all-to-all connectivity of constant strength $w_{TAD}$.
\textbf{b-e)} Depending on the parameter regime, the two-region system can exhibit a subset of four characteristic steady-state epigenetic patterns: \textbf{b)} uniformly closed, marked by high levels of the repressive modification H3K27me3; \textbf{c)} uniformly open, enriched in activating marks such as H3K4me3 and H3K27ac; \textbf{d)} rugged pattern 1 characterised by silenced TAD region; \textbf{e)} rugged pattern 2, in which the TAD region is accessible.}
\label{Figure2}
\end{center}
\end{adjustwidth}
\end{figure}
% , a heterogeneous landscape where the TAD-like region is silenced (high H3K27me3), flanked by active marks in the diagonal region; \textbf{(E)} rugged pattern 2, in which the TAD-like region is transcriptionally active (high H3K4me3 and H3K27ac), surrounded by repressive marks in the diagonal region

To investigate the impact of sirtuin redistribution between chromatin and DNA damage sites, we begin by conducting a stability analysis of the mean-field equations associated with our stochastic model (see Sections~I.2-I.4 of Supplementary Material). We focus on a simplified architecture in which the chromatin domain is partitioned into two distinct regions (Fig.~\ref{Figure2}a): one characterised by diagonal interactions, representing a locally coiled chromatin fibre, and another exhibiting all-to-all interactions with uniform strength, mimicking a TAD-like domain. We denote the interaction strength in the diagonal region by $w_{diag}$, and the contact strength within the TAD region by $w_{TAD}$.

This two-region model, shown in Section~I.4 of Supplementary Material, was selected for two main reasons. First, its simplified geometry significantly reduces the complexity of the governing equations (see Section~I.4 of Supplementary Material), rendering the stability analysis more tractable. Second, our previous work demonstrated that non-local chromatin interactions, such as those in the TAD-like region, facilitate the emergence of heterogeneous epigenetic landscapes \supercite{stepanova2025understanding}. These landscapes exhibit alternating domains of activating and repressive histone modifications, or `rugged' patterns, which define transcriptionally active and silenced genomic regions, thereby influencing cell identity and gene expression. Within this framework (Fig.~\ref{Figure2}a), we identify two types of rugged epigenetic patterns that can arise: \textbf{rugged pattern 1} and \textbf{rugged pattern 2}, illustrated schematically in Figs.~\ref{Figure2}d and e. Rugged pattern 1 is a heterogeneous landscape where the TAD-like region is silenced (high H3K27me3), flanked by active marks in the diagonal region. Rugged pattern 2 is a complementary pattern to rugged pattern 1, with TAD-like region being transcriptionally active (high H3K4me3 and H3K27ac) and surrounded by repressive marks in the diagonal region. In addition to these heterogeneous configurations, the model also admits two uniform epigenetic states: a uniformly \textbf{closed} pattern, enriched in H3K27me3 (Fig.~\ref{Figure2}b), and a uniformly \textbf{open} pattern, marked by high levels of H3K4me3 and H3K27ac (Fig.~\ref{Figure2}c).

\begin{figure}[!htbp]
\begin{adjustwidth}{-0.5cm}{-0.5cm}
\begin{center}
\vspace{-0.3cm}
\includegraphics[width=19cm]{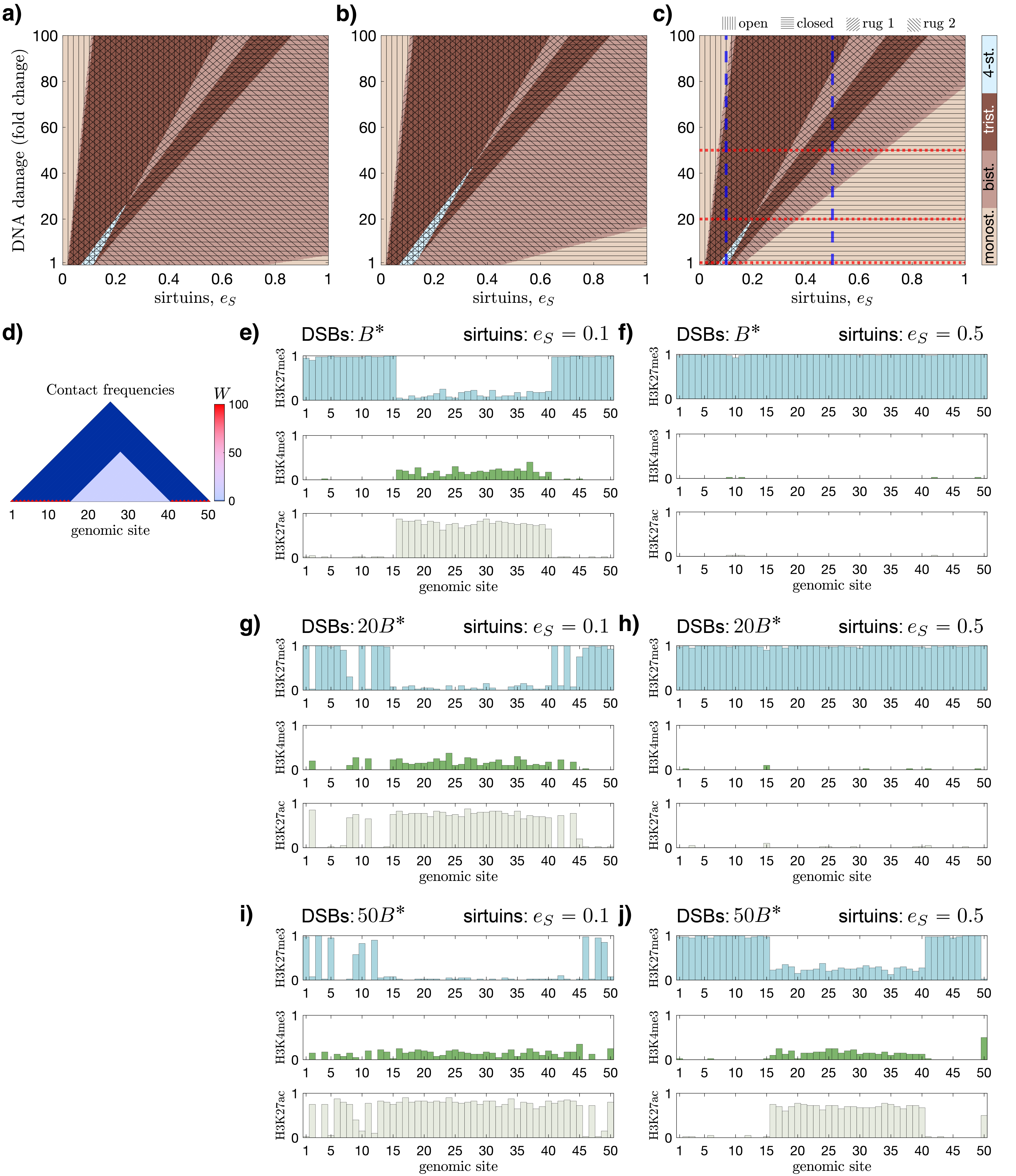}
\caption{\textbf{Epigenetic landscape destabilisation in response to DNA damage in a two-region chromatin model for varying sirtuin concentrations.} \textbf{a-c)} Phase diagrams of the two-region system showing the number and type of stable steady states for variations in sirtuin concentration ($e_S$) and increasing levels of DNA damage (DSBs), relative to the baseline $B^*$. The contact intensity in the diagonal region is fixed at $w_{diag} = 100$, size of the TAD region to $\bar{R} = 25$, while the TAD connectivity varies across panels: \textbf{a)} $w_{TAD} = 100$, \textbf{b)} $w_{TAD} = 50$, \textbf{c)} $w_{TAD} = 10$. The colour bar (rightmost panel) indicates the number of stable steady states, and hatch patterns denote their types: vertical for uniformly open, horizontal for uniformly closed, diagonal ($\nearrow$) for rugged 1, and diagonal ($\searrow$) for rugged 2. The underlying mean-field equations are described in Section~I.4. \textbf{d-j)} Numerical simulations of the stochastic model confirm the predictions of panel \textbf{c} for two parameter regimes (dashed blue lines in \textbf{c}): low sirtuin ($e_S = 0.1$; \textbf{e, g, i}) and high sirtuin ($e_S = 0.5$; \textbf{f, h, j}). At $e_S = 0.1$, \textbf{e)} an initial rugged 2 pattern flattens toward a uniformly open state under \textbf{g)} 20-fold and \textbf{i)} 50-fold increases in DSBs. At $e_S = 0.5$, \textbf{f)} the silenced pattern at baseline DSBs persists at \textbf{h)} 20-fold damage \textbf{j)} but transitions to a rugged state at 50-fold. DSB levels used in simulations are marked by red dotted lines in \textbf{c}. Details of the stochastic simulations are provided in Section~I.2; all other parameters are listed in Supplementary Table~1.}
\label{Figure3}
\end{center}
\end{adjustwidth}
\end{figure}

We performed a parameter sweep of the two-region system to investigate how increasing levels of DNA damage affect the resulting epigenetic patterns. In Figs.~\ref{Figure3}a-c, we present phase diagrams illustrating the range of possible epigenetic states as we vary sirtuin levels and increase the number of DSBs above the baseline value, $B = B^*$ (here, $B$ represents the level of DSBs; details on baseline and pathological DSB levels are provided in Section~\ref{main:methods13}). Two distinct regimes emerge depending on the availability of sirtuin enzymes. At low sirtuin levels (e.g. $e_S = 0.1$, indicated by the left blue line in Fig.~\ref{Figure3}c), and in the presence of baseline DSBs, the system typically exhibits multistability, including heterogeneous rugged epigenetic states. However, as genomic instability intensifies, these rugged landscapes tend to erode into a uniform, open chromatin configuration enriched in activating marks. When sirtuin levels are high (e.g. $e_S = 0.5$ in Fig.~\ref{Figure3}c), chromatin silencing is favoured due to the removal of activating acetylation marks by this enzyme and the system initially adopts a closed chromatin state. However, under elevated DSB conditions, this silenced state is destabilised and replaced by rugged patterns even in sirtuin-rich regimes. Thus, at low sirtuin levels, increasing DSBs drive the system from a multistable regime with heterogeneous patterns toward a monostable, uniformly open chromatin state. In contrast, at higher sirtuin concentrations, an initially stable, silenced chromatin configuration becomes destabilised, giving rise to multistability and the potential emergence of heterogeneous epigenetic landscapes.
 
These findings are consistent with experimental observations reported by Yang et al.\supercite{yang2023loss}, which showed that increased DSBs disrupt pre-existing epigenetic landscapes. Specifically, their results revealed the appearance of new epigenetic marks in previously unmarked regions and a reduction in the levels of modifications initially present, supporting the model's prediction of epigenetic deregulation under DNA damage stress.

We validated the predictions of our steady-state mean-field analysis by performing stochastic simulations of the multi-site model, using the chromatin architecture shown in Fig.~\ref{Figure3}d (corresponding to the parameter values as in Fig.~\ref{Figure3}c). Simulations were performed for both low and high sirtuin concentrations, under progressively increasing levels of DNA damage. At low sirtuin levels ($e_S = 0.1$), the stability analysis predicted that a 20-fold increase in DSBs relative to the baseline level ($B = 20B^*$) is sufficient to destabilise the initial epigenetic landscape (Fig.~\ref{Figure3}c). Consistent with this prediction, comparison of Figs.~\ref{Figure3}e and \ref{Figure3}g (at $B = B^*$ and $B = 20B^*$, respectively) reveals loss of pronounced epigenetic compartmentalisation between the two chromatin regions when $B = 20B^*$, with further erosion toward a uniformly open, H3K27ac-enriched state when $B = 50B^*$ (Fig.~\ref{Figure3}i). By contrast, at high sirtuin levels ($e_S = 0.5$), the closed chromatin pattern predicted at baseline damage (Fig.~\ref{Figure3}f) remains stable under a 20-fold increase in DSBs (Fig.~\ref{Figure3}h), but ultimately destabilises into a heterogeneous state at $B = 50B^*$ (Fig.~\ref{Figure3}j), in agreement with the model analysis (Fig.~\ref{Figure3}c).

We note that the epigenetic configurations observed in the stochastic simulations (Figs.~\ref{Figure3}e-j) are more diverse than the discrete pattern types used in the phase diagram classification (see also Figs.~\ref{Figure2}b-e). This is due to differences in initial conditions: the mean-field analysis was performed for a simple scenario, with homogeneous initial conditions within each region, whereas the stochastic simulations were initialised with randomly distributed epigenetic marks, leading to more varied landscape outcomes.

The results shown in Figs.~\ref{Figure3}a-c were obtained for a chromatin geometry in which the diagonal and TAD-like regions are of equal size (see Fig.~\ref{Figure3}d). Similar trends of epigenetic landscape destabilisation under elevated DSB levels are observed when the size of the TAD region is varied. Specifically, Supplementary Fig.~1 (see Section~II of Supplementary Material) presents the results of a stability analysis for a smaller TAD region (size of TAD region, $\bar{R} = 10$; panels A-C) and a larger one ($\bar{R} = 40$; panels D-F), with the total chromatin domain fixed at $N = 50$ genomic sites. In both cases, the qualitative behaviour remains consistent: at low sirtuin levels, multistable regimes collapse into monostable open chromatin states as DSB levels rise, while at high sirtuin levels, stable closed chromatin is destabilised by increasing DNA damage, leading to the emergence of rugged, heterogeneous patterns.

Our analysis also allows us to predict how epigenetic patterns observed under baseline DNA damage levels could be restored. A consistent trend observed in Figs.~\ref{Figure3}a-c and Supplementary Fig.~1 is that increasing sirtuins above their initial levels can partially re-establish the original epigenetic state, even at elevated DSBs. We tested this prediction through numerical simulations based on the setups shown in Figs.~\ref{Figure3}g, i, and j, where initial patterns were disrupted by DNA damage. Supplementary Fig.~2 shows the resulting epigenetic configurations following sirtuin upregulation. In the case of low sirtuin levels ($e_S = 0.1$), where rugged pattern 2 was lost under 20- and 50-fold increases in DSBs, we were able to recover this pattern by raising the sirtuin concentration to $e_S = 0.2$ (Supplementary Figs.~2a-b). For high sirtuin levels ($e_S = 0.5$), the closed chromatin configuration remained stable under moderate damage ($B = 20B^*$; see Fig.~\ref{Figure3}h), and thus required no additional intervention. However, when DSBs increased to $B = 50B^*$, restoring the silenced state observed at baseline ($B = B^*$; Fig.~\ref{Figure3}f) was possible by further increasing sirtuin levels to $e_S = 0.8$ (compare Fig.~\ref{Figure3}j and Supplementary Fig.~2c). This result is consistent with the mechanism by which DNA double-strand breaks sequester sirtuin enzymes at damage sites; consequently, restoring the original epigenetic patterns under elevated DSBs requires an external supply of additional sirtuins to compensate for their local depletion.

\subsection{Larger domains of non-local chromatin contacts are more robust under DNA damage} \label{main:results2}

We next investigate the role played by the chromatin folding architecture on the robustness of epigenetic landscapes under elevated DNA damage, focusing on whether certain geometric configurations provide higher levels of robustness under local sirtuin depletion. To explore this, we extend the analysis of the two-region system to examine how changes in the size and connectivity of the TAD-like domain affect epigenetic stability in the presence of increased DSBs.

Within our modelling framework, chromatin geometry shapes the epigenetic landscape through two primary mechanisms: enzyme competition between genomic loci (particularly across the two subregions in the two-region system) and feedback from local epigenetic modifications. Enzyme competition is governed by the QSS distributions of enzymes across sites (see Section~I.2), which dynamically allocate a finite pool of enzymes between regions. The feedback mechanisms are encoded in the recruited enzyme activity terms (see Section~\ref{main:methods11} and Eq~(8) in Section~I.2 of Supplementary Material), which depend on the total interaction strength of each site with its neighbours. Due to the simplified structure of our two-region system, these feedback terms become simplified in the mean-field differential equations associated with our model (see Eq~(28) in Section~I.4). In particular, each site in the diagonal region receives only self-feedback, which contributes a constant interaction term of strength $w_{diag}$. In contrast, each site in the TAD-like domain is stimulated by all other sites in the region, with a total strength proportional to $\of{\bar{R} \cdot w_{TAD}}$ due to the all-to-all connectivity within this region.

\begin{figure}[htpb]
\begin{adjustwidth}{-0.5cm}{-0.5cm}
\begin{center}
\vspace{-0.3cm}
\includegraphics[width=19cm]{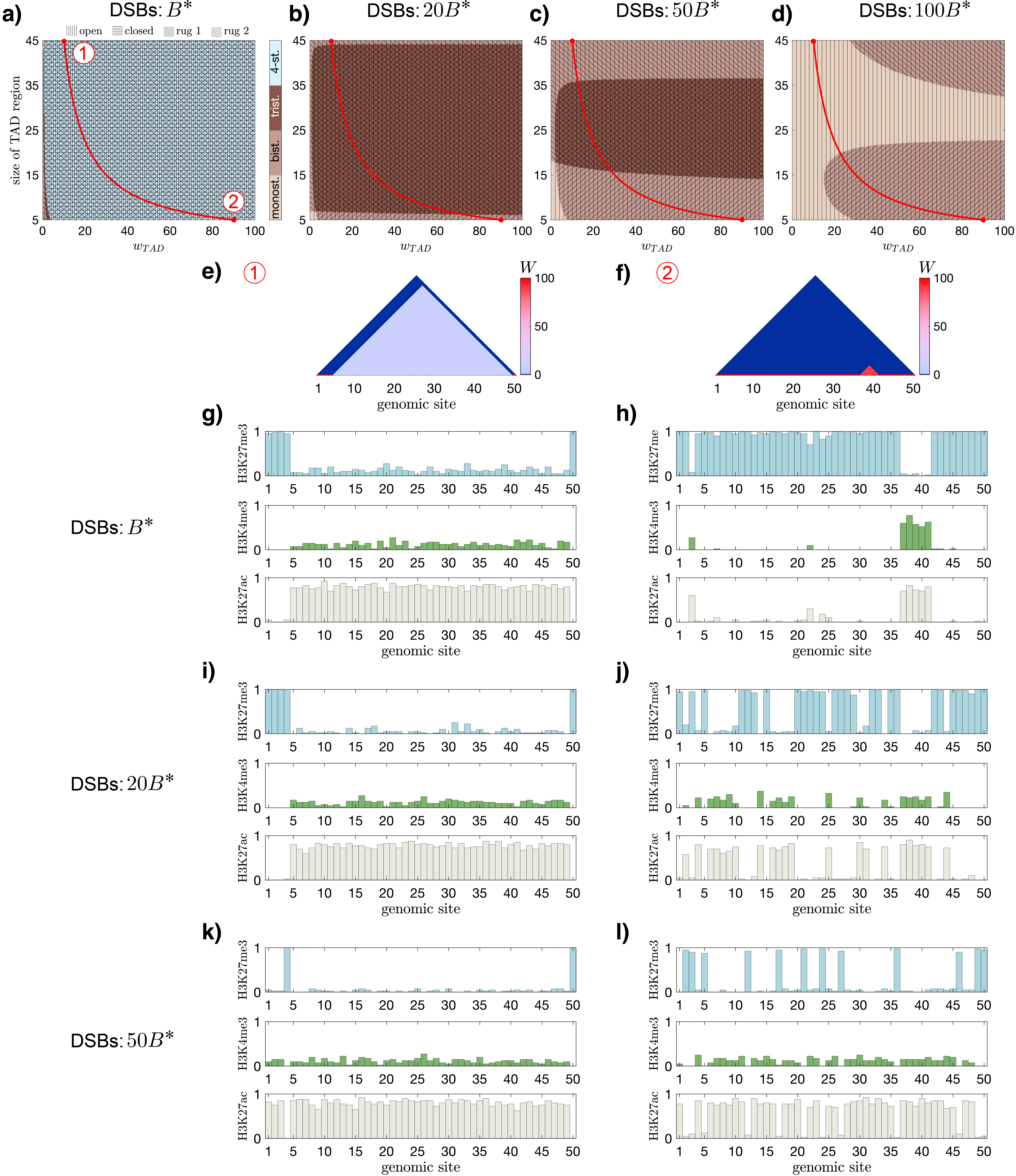}
\caption{\textbf{Geometry-dependent epigenetic stability under increasing DNA damage at low sirtuin concentrations.} \textbf{a-d)} Phase diagrams of the two-region system under low sirtuin availability ($e_S = 0.1$) showing the number and type of stable steady states under varying interaction strength within the TAD-like region ($w_{TAD}$) and its size ($\bar{R}$), for different levels of DNA damage: \textbf{a)} baseline DSB level, $B = B^*$, \textbf{b)} 20-fold, \textbf{c)} 50-fold, and \textbf{d)} 100-fold increases in DSBs. The contact intensity within the diagonal region is fixed at $w_{diag} = 100$, and the total number of genomic sites is set to $N = 50$. The red hyperbolic curve in each panel corresponds to configurations where $\of{\bar{R} \cdot w_{TAD}} = 450$. The colour map and hatching are as in Figs.~\ref{Figure3}a-c. The mean-field equations used are described in Section~I.4. \textbf{e-f)} Two chromatin geometries used in stochastic simulations of the multi-site model highlighted in \textbf{a} ({\protect\circled{1}} and {\protect\circled{2}}). Both geometries satisfy $\of{\bar{R} \cdot w_{TAD}} = 450$, but differ in structure: geometry 1 has a large TAD region with weaker internal contacts (\textbf{e}: $\bar{R} = 45$, $w_{TAD} = 10$), while geometry 2 features a compact TAD region with stronger internal contacts (\textbf{f}: $\bar{R} = 5$, $w_{TAD} = 90$). \textbf{g, i, k)} Final epigenetic landscapes from stochastic simulations for geometry 1 (\textbf{e}) under \textbf{g)} baseline DSBs, \textbf{i)} 20-fold, and \textbf{k)} 50-fold DSB increases. Corresponding results for geometry 2 (\textbf{f}) are shown in \textbf{h, j, l)}. Simulation details are provided in Section~I.2; all other parameters are listed in Supplementary Table~1.}
\label{Figure4}
\end{center}
\end{adjustwidth}
\end{figure}

Therefore, to isolate the effect of chromatin geometry via enzyme competition alone, we fix $w_{diag}$ and compare chromatin geometries in which the total strength of the feedback remains constant (i.e. $\of{\bar{R} \cdot w_{TAD}}$ is held fixed). This approach allows us to compare the epigenetic dynamics of chromatin architectures with different TAD sizes but equivalent total feedback strength, thereby attributing any observed differences solely to the redistribution of enzymes under damage-induced stress.

Given that the destabilisation dynamics of the epigenetic landscape under increased DSBs differ between low and high sirtuin availability (see Fig.~\ref{Figure3}), we analyse these two regimes separately. Results for low sirtuin concentrations are shown in Fig.~\ref{Figure4}, while Supplementary Fig.~3 presents the corresponding analysis under high sirtuin conditions. Consistent with our earlier findings (Fig.~\ref{Figure3}, Supplementary Fig.~1), elevated levels of DSBs reduce the number of stable steady states across all chromatin geometries when sirtuin levels are low (Figs.~\ref{Figure4}a-d). However, this loss of stability is not uniform across the chromatin folding configurations represented in these phase diagrams. To explore this further, we compare two specific geometries, marked by {\protect\circled{1}} and {\protect\circled{2}} in Fig.~\ref{Figure4}a, that share the same total feedback strength ($\bar{R} \cdot w_{TAD} = \text{const}$), but differ in structure: geometry 1 features a large TAD region with weaker internal interactions, while geometry 2 has a compact TAD region with stronger interlocus connectivity (see schematics in Figs.~\ref{Figure4}e-f).

Our previous modelling work suggests that chromatin domains with non-local interactions often exhibit acetylation patterns concentrated around regions with high contact frequency, a configuration we classify as rugged 2 (Fig.~\ref{Figure2}e) \supercite{stepanova2025understanding}. At baseline DNA damage ($B = B^*$) with a low level of sirtuins, both geometries exhibit this rugged 2 pattern (Fig.~\ref{Figure4}a). However, as DNA damage levels increase, the rugged 2 landscape in geometry 2 begins to erode and eventually disappears (Figs.~\ref{Figure4}b-c), while geometry 1 remains more resilient. Only at extreme damage levels (100-fold increase in DSBs) are both geometries predicted to exhibit substantial degradation of their epigenetic landscapes (Fig.~\ref{Figure4}d). These results from the mean-field model suggest that chromatin architectures with larger domains of non-local contacts are more robust to elevated DNA damage than those with smaller domains of non-local contacts.

Since the mean-field analysis assumes a simplified scenario, we also performed stochastic simulations for the two chromatin geometries. Figs.~\ref{Figure4}g, i, and k show the final epigenetic landscapes for geometry 1 under increasing DNA damage: baseline ($B = B^*$), 20-fold, and 50-fold DSB increase, respectively. As predicted by the mean-field analysis, the rugged 2 pattern persists under moderate DNA damage ($B = 20B^*$) but eventually collapses into a near-uniform, acetylation-dominated landscape at $B = 50B^*$. In contrast, geometry 2 is markedly more vulnerable: a 20-fold increase in DSBs is sufficient to disrupt the original pattern (compare Figs.~\ref{Figure4}h and j), and further damage pushes the system toward a globally open chromatin state (Fig.~\ref{Figure4}l). Taken together, these findings suggest that larger domains of non-local interactions confer greater robustness to sirtuin depletion, even when internal contact frequencies are relatively low, likely because the broader interaction network distributes the burden of enzymatic competition more effectively under stress.

We observe similar trends when sirtuin levels are high (see Supplementary Fig.~3). In this parameter range, our earlier analysis (Fig.~\ref{Figure3} and Supplementary Fig.~1) indicates that initially monostable, silenced chromatin at baseline DSB levels becomes more permissive under elevated levels of DNA damage, often through the emergence of additional stable equilibria. Extending this analysis to variations in chromatin geometry confirms this general trend but also reveals that the stability of the silenced epigenetic landscape depends strongly on the size and structure of the TAD-like region. Specifically, smaller TAD regions (for example, geometry 2) are more susceptible to destabilisation, losing their silenced state at lower levels of DNA damage than larger domains with non-local connectivity (e.g. geometry 1), as seen in Supplementary Figs.~3a-d. For instance, a 20-fold increase in DSBs is sufficient to disrupt the closed chromatin configuration in geometry 2, while geometry 1 remains stable under these conditions (Supplementary Fig.~3b). These predictions were validated through stochastic simulations (Supplementary Figs.~3e-l), which show that geometry 1 maintains the silenced epigenetic pattern at $B = 20B^*$, while geometry 2 collapses into a disordered, less stable state (compare Supplementary Figs.~3i and j). 

Taken together, the above results suggest that the enhanced robustness of larger TAD-like domains under elevated DNA damage is a general property of the system. This conclusion holds regardless of the initial sirtuin availability, or equivalently, the initial epigenetic configuration (i.e. under baseline levels of DSBs), since sirtuin levels directly shape the underlying landscape. By fixing the total feedback strength independent of the chromatin geometries ($\bar{R} \cdot w_{TAD} = \text{const}$), we isolated the specific contribution of chromatin geometry arising from enzyme binding competition, independent of its influence on feedback strength. The observed stability of larger TAD domains thus appears to stem from our modelling assumption that enzyme binding does not depend on contact intensity. As a result, larger domains offer more genomic sites where the correct modifying enzymes can bind. Once present, the feedback mechanisms can amplify local epigenetic signals, helping to sustain the intended chromatin state even under DSB-induced sirtuin depletion.

\begin{figure}[htbp]
\begin{adjustwidth}{-0.5cm}{-0.5cm}
\begin{center}
\vspace{-0.8cm}
\includegraphics[width=19cm]{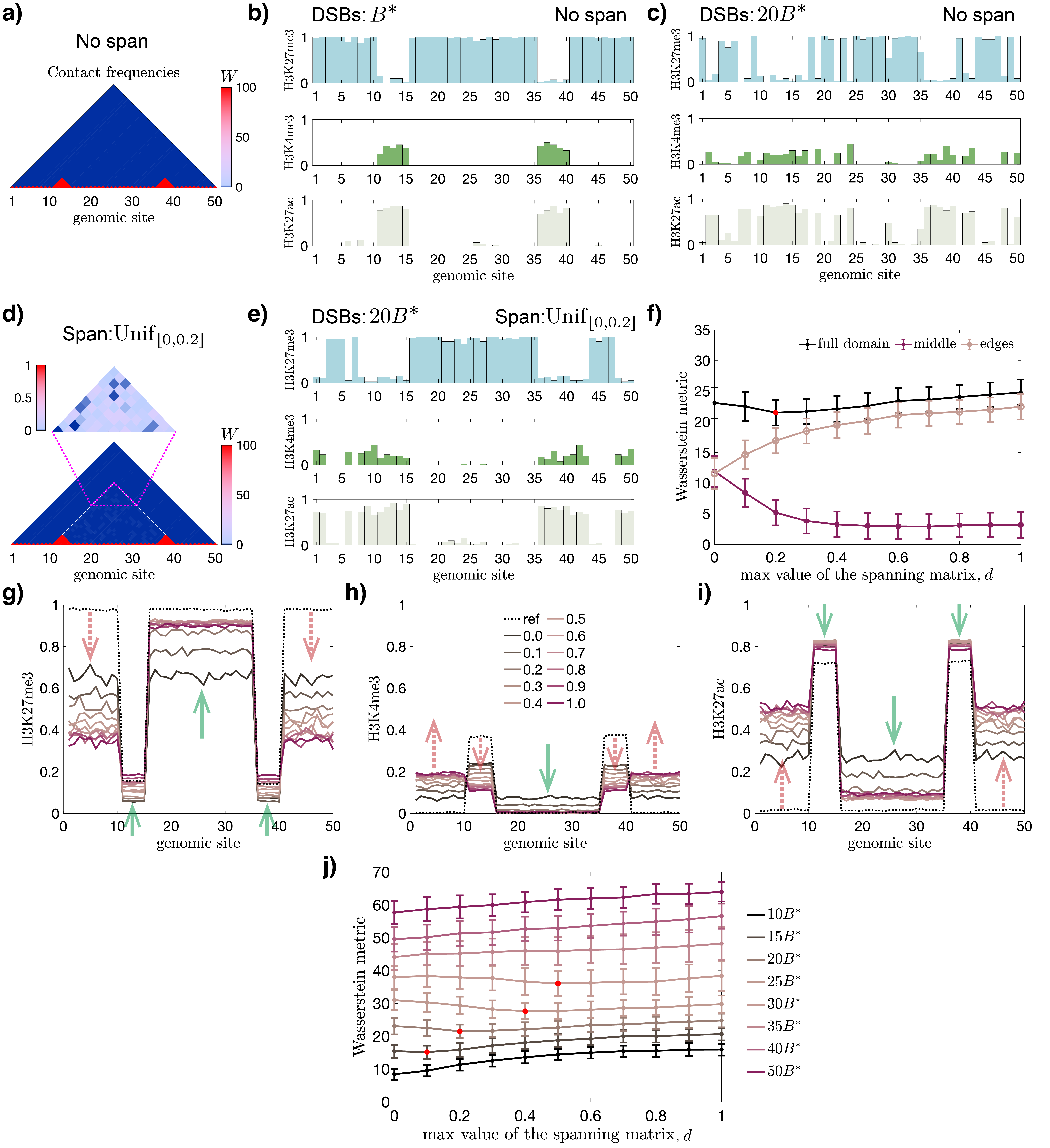}
\caption{\textbf{Stochastic simulations of the multi-site model for a chromatin geometry with two non-local interaction hubs separated by a locally coiled region.} \textbf{a)} Chromatin contact map without a spanning domain. \textbf{b-c)} Final epigenetic landscapes for configuration \textbf{a} under \textbf{b)} baseline DNA damage, $B = B^*$ and \textbf{c)} a 20-fold increase in DSBs. \textbf{d)} Chromatin geometry with an added low-intensity spanning domain (contacts sampled from $\unif_{[0,0.2]}$), outlined by a dashed white line. \textbf{e)} Final epigenetic pattern for geometry \textbf{d} under $B = 20B^*$, partially restoring the baseline pattern from \textbf{b}. \textbf{f)} Wasserstein distance between the epigenetic patterns under the baseline amount of DSBs, $B = B^*$ and a 20-fold increase, $B = 20B^*$, for increasing maximum contact value $d$ of the spanning matrix. The metrics are computed for the full domain (sites 1-50, black), the central region (11-40, magenta), and the outer flanks (1-10 and 41-50, beige). We used the Python Optimal Transport package for the numerical computation of the Wasserstein metric \supercite{flamary2021pot}. Error bars indicate standard deviation calculated over 500 realisations; the red dot marks the optimal $d = 0.2$ minimising divergence from the baseline. \textbf{g-i)} Mean epigenetic profiles (averaged over 500 realisations) for increasing $d$ under $B = 20B^*$ for \textbf{g)} H3K27me3 ($\mathbf{p_I}$), \textbf{h)} H3K4me3 ($\mathbf{p_A}$), and \textbf{i)} H3K27ac ($\mathbf{q}$). Dashed black lines show baseline patterns at $B = B^*$ with no spanning, while coloured solid lines show results for increasing $d$ (legend in \textbf{h}). Green arrows highlight regions where increased $d$ improves recovery; red dashed arrows indicate worsening at the flanks. \textbf{j)} Wasserstein distances for the entire chromatin domain under varying DSB levels (from $B = 10B^*$ to $B = 50B^*$) and increasing $d$. Red circles mark local minima, indicating optimal recovery at intermediate $d$ values. All metrics computed from 500 realisations per condition; error bars represent standard deviation.}
\label{Figure5}
\end{center}
\end{adjustwidth}
\end{figure}

Motivated by these predictions, we tested whether introducing low-intensity contacts around specific regions, particularly those with functional significance for epigenetic regulation, could improve robustness under elevated DNA damage. To this end, we considered a chromatin folding configuration in which two small domains of non-local interactions are separated by a segment of locally coiled chromatin (Fig.~\ref{Figure5}a). This geometry could represent a genomic region encompassing an enhancer and a promoter, as they are typically characterised by genomic sites with non-local interactions separated from each other by a stretch of chromatin with more localised connections. The baseline epigenetic landscape under normal DSB levels (Fig.~\ref{Figure5}b) shows high levels of activating marks within these contact-rich domains, flanked by silenced chromatin. However, upon increasing the number of DSBs, this pattern is destabilised (see Fig.~\ref{Figure5}c for a representative realisation with $B = 20B^*$).

We then tested whether the original pattern could be partially recovered by introducing low-frequency interactions in a `spanning domain' connecting the two contact-rich regions. A representative spanning matrix is shown in Fig.~\ref{Figure5}d, where the spanning region is outlined by a white dashed line for clarity. Contact values within this region were sampled from a uniform distribution, $\unif_{[0,d]}$, and the maximum value $d$ was systematically increased to determine whether there exists a level of spanning interactions that can restore the original epigenetic pattern. An example of partial recovery using $d = 0.2$ is shown in Fig.~\ref{Figure5}e.

We performed 500 stochastic simulations under a 20-fold increase in DSBs, progressively varying the spanning contact strength from $d = 0$ to $d = 1$ in steps of size $0.1$. At each realisation, the initial epigenetic state and the spanning matrix were sampled independently, ensuring robust conclusions regardless of the effects of local perturbations. To quantify pattern recovery, we used the Wasserstein distance, $W(X_1,X_2)$, a metric that captures the approximate `cost' associated with transforming one empirical distribution into another \supercite{flamary2021pot}. Here, the epigenetic landscape, i.e. the vector $\of{\mathbf{p_I}, \mathbf{p_A}, \mathbf{q}}$, was normalised to yield empirical distributions. The reference distribution $X_1$ was computed from the average landscape over 500 simulations at baseline DSBs without spanning (Figs.~\ref{Figure5}a-b), while $X_2$ was computed for individual realisations with $d>0$ spanning and $B = 20B^*$ DSBs.

Fig.~\ref{Figure5}f shows how Wasserstein distance metrics, calculated over the full chromatin domain (black), the central spanning region (sites $11-40$, dark magenta), and the outer edges (sites $1-10$ and $41-50$, beige), vary with increasing $d$. We calculated the Wasserstein metrics separately for the three regions, as changes in epigenetic patterns differ between the central spanning region and the domain flanks. In particular, as $d$ increases, the epigenetic pattern within the spanning domain becomes more similar to the baseline landscape, whereas the pattern in the outer regions is more divergent (see Figs.~\ref{Figure5}g-i). These opposing trends result in a local minimum in the global Wasserstein distance, identifying the optimal balance between stabilisation in the spanning region and disruption at the edges. For example, for DSB levels with $B = 20B^*$, $d = 0.2$ yields the optimal recovery (highlighted by a red circle in Fig.~\ref{Figure5}f), with a representative restored pattern shown in Fig.~\ref{Figure5}e.

This trade-off is further illustrated in Figs.~\ref{Figure5}g-i, which show average epigenetic profiles for different values of $d$. Within the spanning region, increasing $d$ generally improves pattern recovery (green arrows), except for the H3K4me3 mark. However, increasing $d$ causes further disruption at the edges (red dashed arrows), confirming the spatial specificity of the spanning effect. Nevertheless, intermediate values of $d$ can partially restore the overall epigenetic landscape.

To assess whether this mechanism generalises across varying damage levels, we extended our analysis to DSB levels in the range $10B^* \leq B \leq 50B^*$. As shown in Fig.~\ref{Figure5}j, at low levels of damage ($B = 10B^*$), the original pattern remains relatively intact, and spanning introduces unnecessary disruption. Conversely, at high damage levels ($B \geq 35B^*$), the pattern is so disrupted that the spanning domain is ineffective. However, for intermediate damage levels ($15B^* \leq B \leq 30B^*$), the spanning domain aids recovery, as indicated by local minima in the Wasserstein curves (red circles). Supplementary Fig.~4 shows individual realisations of the epigenetic landscape before and after spanning for $B = 15B^*$ and $B = 25B^*$ DSBs, highlighting this partial restoration.

These results suggest that increasing the spatial extent of chromatin domains with non-local interactions, even with low-intensity contacts, can enhance the robustness of epigenetic patterns under moderate levels of DNA damage. In this case, the total feedback strength varied, implying that the stabilising effect can arise from the combined contributions of competition for enzyme binding and feedback from existing chromatin modifications. This finding suggests that experimentally observed configurations, such as loops between chromatin regulatory elements (promoters, enhancers, insulators) embedded within broader regions of elevated contact frequency (e.g. \supercite{hsieh2022enhancer,aljahani2022analysis,lee2022characterizing,costantino2020cohesin}), may play a functional role in maintaining epigenetic stability under increased DNA damage.

\begin{figure}[htbp]
\begin{adjustwidth}{-0.5cm}{-0.5cm}
\begin{center}
\includegraphics[width=19cm]{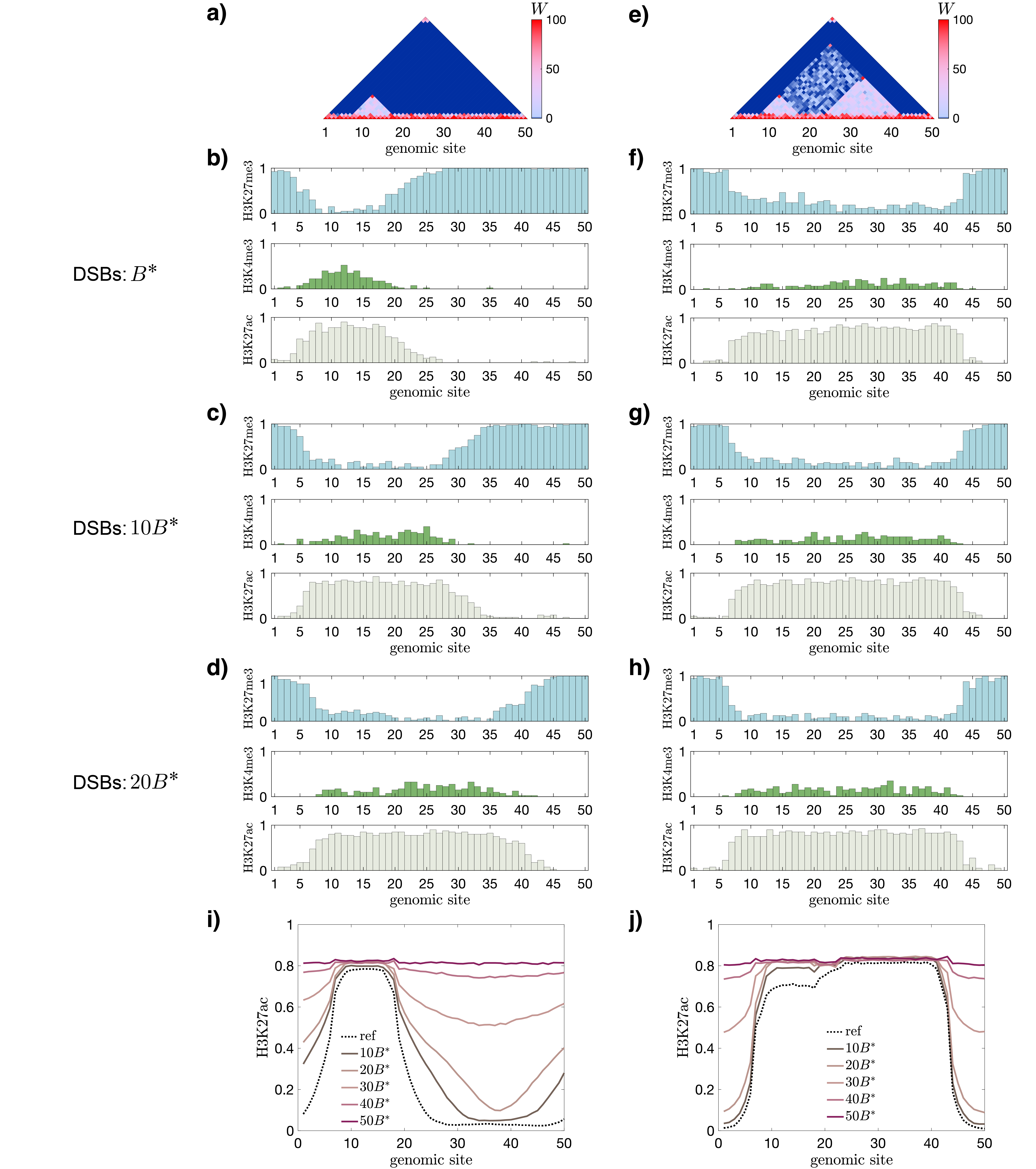}
\caption{\textbf{Stochastic simulations of epigenetic landscapes for more realistic chromatin architectures.} \textbf{a-d)} Representative epigenetic patterns for a chromatin configuration featuring a locally coiled DNA polymer with a single small TAD domain shown in \textbf{a)}. The resulting landscapes are shown for \textbf{b)} baseline DSB levels, $B = B^*$, and for elevated damage: \textbf{c)} 10-fold and \textbf{d)} 20-fold increases. Contact values within the TAD region are sampled from $\unif_{[0,40]}$, while diagonal elements are sampled from $\unif_{[60,100]}$. \textbf{e-h)} Representative epigenetic patterns for a chromatin architecture with two small TAD domains embedded within a larger region of low-intensity contacts, shown in \textbf{e)}. Corresponding epigenetic landscapes are shown for \textbf{f)} baseline DSBs, \textbf{g)} 10-fold, and \textbf{h)} 20-fold DSB increases. Contact values within the TAD regions are sampled from $\unif_{[0,40]}$, the surrounding embedding region from $\unif_{[0,2]}$, and diagonal values from $\unif_{[60,100]}$. \textbf{i, j)} Averaged H3K27ac patterns for the chromatin geometries in \textbf{a} and \textbf{e}, respectively, under increasing DSB levels. Dashed black lines show the average landscape at baseline ($B = B^*$), while solid lines correspond to elevated DSBs ($10B^* \leq B \leq 50B^*$; see colour legend). Results are averaged over 500 stochastic realisations for each condition. Simulation details are provided in Section~I.2; all parameters are listed in Supplementary Table~1.}
\label{Figure6}
\end{center}
\end{adjustwidth}
\end{figure}

\subsection{Robustness of larger TAD domains in more complex chromatin foldings} \label{main:results3}

We now investigate whether the observed improvements in epigenetic robustness extend to more realistic folding configurations. To this end, we examine two contact-frequency matrices, shown in Figs.~\ref{Figure6}a and e. Both matrices feature high-intensity interactions along the main diagonal, representing local chromatin coiling. In addition to this `low-order' structure, the first configuration (Fig.~\ref{Figure6}a) contains a single, small, TAD-like domain, enriched in non-local contacts, whereas the second (Fig.~\ref{Figure6}e) includes two such domains embedded within a broader region of low-intensity interactions. To account for biological variability, random noise is added to the interaction strengths in both geometries (see Fig.~\ref{Figure6} caption for details). We note that these folding patterns are qualitatively similar to structures found in experimental chromatin conformation data (e.g. see \supercite{rao20143d,szabo2019principles,sun2024mapping,mccord2020chromosome,costantino2020cohesin,tavallaee2024mapping}).

Figs.~\ref{Figure6}b-d show individual realisations of the epigenetic landscape for the first chromatin architecture (panel A) under progressively increasing levels of DSBs. Initially, activating epigenetic marks are confined to the non-local TAD region (Fig.~\ref{Figure6}b). However, as DSB levels increase, these modifications expand into adjacent regions, previously dominated by repressive H3K27me3 marks. This pattern is reflected in the average acetylation profiles over 500 stochastic simulations (Fig.~\ref{Figure6}i), which reveal early spreading of activating marks even under modest damage (e.g. $10B^*$), suggesting a potential loss of repression and increased transcriptional accessibility outside the original TAD.

By contrast, the second geometry, with two embedded TADs (Fig.~\ref{Figure6}e), exhibits greater robustness to increasing DSB levels, maintaining compartmentalisation of activating and repressive marks up to a critical threshold of $B \approx 30B^*$ (Figs.~\ref{Figure6}f-h, j). In this case, the boundaries of the TAD domains coincide with clear transitions in the epigenetic state, demonstrating sharper and more persistent domain-specific epigenetic signatures.

These findings are not specific to the contact matrices shown in Fig.~\ref{Figure6}. Supplementary Figs.~5-6 replicate the analysis with variations in the strength of non-local contacts. In all cases, we find that smaller domains are more vulnerable to epigenetic destabilisation, even at low DSB levels, whereas larger domains maintain their integrity up to higher levels of DNA damage.

While a broader range of chromatin architectures could be explored, the folding patterns shown in Fig.~\ref{Figure6} and Supplementary Figs.~5-6 are among the most commonly encountered in experimental studies. Excluding purely locally coiled configurations (which rarely exhibit domain-specific epigenetic signatures), our results suggest that variations around the motifs presented here are likely to yield similar outcomes within the proposed modelling framework.

Taken together, these findings support the conclusion that larger epigenetic domains, and more generally, chromatin configurations featuring non-local contacts, provide enhanced stability of epigenetic patterns in response to increased levels of DSBs. This highlights a potentially important functional role of 3D genome organisation in maintaining epigenetic robustness under DNA damage.

\section{Discussion} \label{main:discussion}

In this study, we have investigated how the local biochemical environment and chromatin folding influence the stability of epigenetic landscapes under local depletion of sirtuins, histone deacetylases that also participate in DNA repair, at sites of DSBs. Building on our previous modelling work \supercite{stepanova2025understanding}, which integrates epigenetic regulation within the three-dimensional chromatin structure, we extended this framework to account for the relocalisation of sirtuins from histone tails to DNA damage sites. This enabled us to simulate epigenetic landscape destabilisation under increased DSB levels and to explore how local concentrations of histone-modifying enzymes and chromatin architecture affect the erosion dynamics of these patterns. Our findings offer mechanistic insights into the structural principles underlying epigenetic robustness in the face of elevated DNA damage.

Our model reproduces the experimental observations reported in Yang et al.\supercite{yang2023loss}, which show that sirtuin sequestration to DSB sites leads to erosion of epigenetic patterns. Specifically, they observed that epigenetic erosion typically involves a reduction in pre-existing modifications and an increase in marks previously absent, following non-mutagenic, externally induced DSBs (ICE system \supercite{yang2023loss}). Our simulations confirm this trend: under elevated DSB conditions, activating modifications (H3K4me3 and H3K27ac) tend to increase due to the partial relocation of sirtuins (H3K27ac deacetylases) to DSB sites, while repressive marks such as H3K27me3 decrease. Consistent with this mechanism, we also find that supplementing sirtuins can restore the original epigenetic patterns disrupted by high levels of DSBs.

However, epigenetic destabilisation varies with the availability of sirtuins. At low concentrations of this enzyme, our model predicts that initially heterogeneous patterns, comprising alternating activating and repressive marks, become increasingly enriched in activating marks as DSB levels rise. At high levels of DNA damage, the chromatin region converges to a uniform pattern dominated by activating marks. In contrast, under intermediate-to-high initial sirtuin levels, initially silenced regions become more permissive due to partial replacement of repressive marks by their antagonistic activating modifications. Thus, initially homogeneous repressive landscapes become heterogeneous. These contrasting effects highlight that epigenetic erosion is not uniform; rather, it depends critically on the biochemical context. Different types of landscapes degrade differently: rugged configurations may lose structure entirely, while compacted domains may fragment into heterogeneous patterns.

Chromatin geometry also plays a key role in determining the robustness of epigenetic patterns under sirtuin depletion. Analysis of our model and stochastic simulations reveal that larger domains with non-local interlocus contacts, even at low interaction strengths, exhibit greater robustness against high DSB levels. This suggests that chromatin folding, particularly higher-order structures such as loop domains, may act as protective mechanisms against perturbations. Furthermore, our simulations show that even weak long-range interactions between contact-rich domains can partially rescue disrupted epigenetic profiles. This robustness arises from competitive enzyme binding and the propagation of epigenetic marks through feedback loops facilitated by chromatin folding. In our model, larger domains of non-local interactions are less susceptible to sirtuin relocation, as they can redistribute the remaining enzyme within the domain and maintain epigenetic marks through feedback.

These findings align with experimental results from Yang et al.\supercite{yang2023loss}, which report the formation of new interlocus contacts in ICE-treated cells with induced DSBs near the Hoxa genes, effectively expanding the domain of non-local contacts. We propose that these new contacts may not merely be by-products of damage but could represent compensatory responses aimed at restoring epigenetic stability. In our model, such interactions function like `support scaffolds', enabling the re-establishment of feedback loops between reader-writer enzymes, even when the local contact network is compromised due to sirtuin depletion.

Our model predictions regarding the increased robustness of chromatin regions with larger domains of non-local contacts could be potentially tested experimentally by comparing the stability of loop domains of different sizes under induced DNA damage or localised sirtuin depletion. Such experiments would require that the compared domains share similar biochemical environments, since, as our results show, the erosion of epigenetic patterns depends on local concentrations of histone-modifying enzymes (and likely other nuclear components not included in our model). An alternative approach could involve artificially increasing contact frequencies between specific loci using dCas9-mediated looping \supercite{hao2017programmable}. Comparing chromatin domains before and after loop induction under elevated DSB levels would allow experimental assessment of whether such loops enhance resistance to epigenetic erosion.

A relevant simplification in our modelling framework is the assumption that enzyme binding rates are independent of chromatin conformation. If enzyme binding is hindered in regions with more intense chromatin contacts, this can potentially affect our results regarding the destabilisation of epigenetic patterns. Nonetheless, we do not expect this limitation to alter our conclusions regarding the added robustness provided by spanning matrices embedding critical epigenetic domains, as the contact frequencies considered in these regions were relatively low compared to overall genome-wide interaction values.

We also note that our current model assumes a static contact frequency matrix defining chromatin architecture. However, histone modifications are known to influence chromatin structure, suggesting that significant damage-induced epigenetic changes may in turn reshape chromatin folding. This feedback could lead to permanent erosion of the original epigenetic profile, even if enzyme levels are later restored. To address this possibility, future work should incorporate reciprocal interactions between the epigenetic landscape and local chromatin organisation.

Overall, our work provides a flexible tool for exploring the mechanisms driving epigenetic landscape erosion. Our results emphasise the importance of chromatin geometry in maintaining epigenetic stability, supporting the view that chromatin folding is not merely a passive context for epigenetic regulation but an active player in the dynamic preservation of epigenetic signatures, and thus cellular identity, under both internal and external perturbations.

\section{Materials and methods} \label{main:methods1}

\subsection{Epigenetic regulation model} \label{main:methods11}

We recently developed a stochastic model for epigenetic regulation that integrates the dynamics of epigenetic mark addition and removal along a chromatin fibre with the spatial architecture of chromatin folding (an example of chromatin conformation is shown in Fig.~\ref{Figure1}a). This model considers epigenetic modifications at the tails of histone H3, at residues 4 and 27. Trimethylation of lysine 27 of this histone type (H3K27me3) is generally considered to be a repressive mark which leads to chromatin silencing. By contrast, the addition of an acetyl group to the same residue (H3K27ac) favours chromatin accessibility and, thus, gene expression. We also consider the activating mark H3K4me3, trimethylation of lysine 4 of histone H3. In a very general way, acetylation is typically associated with chromatin activation, while the effects of methylation of histone tails are context-dependent. The processes of depositing or removing epigenetic marks depend on specific enzymes. Thus, our core model focuses on six enzymatic reactions for adding and erasing epigenetic marks, each reaction being mediated by a specific class of enzymes. The reactions of interest are depicted in Fig.~\ref{Figure1}b. More information can be found in Supplementary Material, our previous work \supercite{stepanova2025understanding}, and references therein. 

The deposition of new epigenetic modifications is influenced by existing marks in the surrounding chromatin environment. Modified lysine residues can recruit reader-writer enzymes which can recognise (or `read') these marks. This allosterically enhances the activity of their writer domains, facilitating the addition of new modifications. In particular, H3K27 trimethylation (H3K27me3) reinforces deposition of the same marks and inhibits the addition of its antagonistic mark, H3K4me3. Conversely, H3K4me3 suppresses the trimethylation of H3K27 in its vicinity.
Additionally, the two activating marks, H3K4me3 and H3K27ac, are known to promote each other’s deposition. Our model incorporates these interactions by making the enzymatic rate constants for mark addition dependent on the surrounding epigenetic landscape \supercite{stepanova2025understanding}. When chromatin is folded, bringing distant genomic loci into close spatial proximity, these feedback mechanisms can extend beyond local interactions. Thus, in our model, the rate functions for epigenetic mark deposition account for the cumulative influence of modifications across all genomic loci, weighted by their interaction intensity due to chromatin folding.

Chromatin folding can be quantified using chromatin conformation capture (CCC) techniques such as genome-wide Hi-C \supercite{lieberman2009comprehensive} or a finer resolution alternative Micro Capture-C \supercite{hamley2023determining}. Both approaches generate contact frequency maps representing pairwise interaction intensities between genomic loci. The resolution, or length, of each genomic site in our model is considered uniform and corresponds to the characteristic resolution of CCC techniques, typically ranging from $200$ to $5000$ base pairs ($0.2-5$ kb). The contact frequency map can be expressed as a symmetric matrix, $W = \uf{w_{ij}}$, where each element, $w_{ij}$, reflects the relative frequency of spatial interactions between genomic sites $i$ and $j$ due to chromatin folding (Fig.~\ref{Figure1}a). %While we refer to these values as contact frequencies, they are dimensionless and do not directly correspond to any physical quantity. 
The magnitude of these contact frequencies varies significantly across CCC methods, resolutions, and sequencing depths. Therefore, we assume that the essential information captured in $W$ lies in the relative differences between low- and high-frequency interactions. Although epigenetic modifications can influence chromatin folding, we assume that these structural changes occur on a slower timescale than the dynamics of epigenetic marks. As a result, we treat $W$ as constant throughout our simulations.

In our previous work \supercite{stepanova2025understanding}, we simplified the original system of enzymatic reactions (Fig.~\ref{Figure1}b) using a quasi-steady-state approximation. This reduction is valid when histone-modifying enzymes are in low abundance relative to the number of lysine residues, an assumption supported by experimental data and previous modelling studies \supercite{stafford2018multiple,owen2023design}. In such cases, the model can be coarse-grained to track the dynamics of three key epigenetic modifications at each genomic site (Fig.~\ref{Figure1}c). This simplification significantly reduces computational complexity, enabling a more comprehensive analysis of how model parameters and chromatin folding influence the formation of heterogeneous epigenetic landscapes \supercite{stepanova2025understanding}.

One key finding, supported by our previous study, is that chromatin subregions with non-local interactions promote the emergence of heterogeneous epigenetic profiles \supercite{stepanova2025understanding}. A similar pattern, generated by our model, is shown in Fig.~\ref{Figure1}c, which corresponds to the chromatin geometry depicted in Fig.~\ref{Figure1}a. These heterogeneous epigenetic landscapes play a crucial role in defining cell identity, as they regulate gene expression by silencing regions enriched in repressive marks while making transcriptionally active regions accessible through activating epigenetic modifications.

A detailed model description, derivation of the QSS assumption and its analysis can be found in Supplementary Material and our recent work \supercite{stepanova2025understanding}.

\subsection{Extended model accounting for sirtuin redistribution to DNA repair sites} \label{main:methods12}

We now use our model to investigate the extent to which genomic instability, marked by an increased number of DSBs, can destabilise epigenetic patterns and drive critical changes in cell identity. Specifically, we build on the hypothesis proposed by Yang et al. \supercite{yang2023loss}, which suggests that sequestration of histone-modifying enzymes, such as sirtuins, to DNA repair sites leads to the erosion of epigenetic landscapes. Sirtuins, a family of enzymes that regulate the levels of the activating mark H3K27ac, also play a key role in DSB repair. Yang et al. demonstrated that when these enzymes are recruited to repair sites, the cell’s epigenetic state is altered, contributing to accelerated ageing \supercite{yang2023loss}. We extend our model to account for the recruitment of sirtuins to DSBs, by incorporating the following reaction into our original framework (Fig.~\ref{Figure1}b):

\begin{align} \label{eq_sirt_dsb}
& B + E_S \stacksymbol{\rightleftarrows}{\textcolor{burgundy}{\bar{k}_{1}}}{\textcolor{burgundy}{\bar{k}_{-1}}} C_{SB}.
\end{align}
\noindent 
Here, $E_S$ represents the level of free sirtuins, while the dimensionless parameter $B$ represents the number of DSBs. The levels of DSB sites with bound sirtuin are denoted by $C_{SB}$. The rate constants $\bar{k}_{1}$ and $\bar{k}_{-1}$ determine the timescales of sirtuin recruitment to DSBs and their subsequent release. For simplicity, we assume a constant level of DSBs during model simulations. Thus, Eq~\eqref{eq_sirt_dsb} provides a straightforward way to account for the redistribution of sirtuins from their usual role in removing acetyl groups from histone tails to DNA repair sites.

We use the extended model (Fig.~\ref{Figure1}b) to investigate how increased levels of DSBs affect the epigenetic landscape. Specifically, we seek to determine whether our model can reproduce the experimental findings of Yang et al., in which high levels of DNA damage lead to epigenetic erosion \supercite{yang2023loss}. We also assess the role of chromatin conformations in preserving the original epigenetic landscape when it is subject to elevated levels of genomic instability.

Following the approach introduced in our previous work \supercite{stepanova2025understanding}, we simplify the extended model (Fig.~\ref{Figure1}b) using a timescale separation assumption. As before, this assumption is based on the low levels of histone-modifying enzymes relative to the large numbers of histone tails to which they can bind. In particular, we assume that enzyme complexes (fast variables) rapidly reach their quasi-steady-state (QSS) distribution while epigenetic modifications (slow variables) evolve over a longer timescale. Under these assumptions, the model dynamics are dominated by the dynamics of epigenetic modifications, with enzyme complexes sampled from their QSS distributions. Similar stochastic model reduction techniques have been employed in other works \supercite{ball2006asymptotic,folguera2019multiscale,alarcon2021}.

In our previous work, we demonstrated that the QSS distribution of enzyme complexes follows a multinomial distribution, whose probabilities describe the likelihood of an enzyme being bound to a genomic site or remaining unbound. In the extended model, the QSS probabilities are unchanged from those in reference\supercite{stepanova2025understanding}, except for the sirtuins, which can also bind to DNA damage repair sites. A full derivation of the QSS approximation can be found in reference\supercite{stepanova2025understanding}; a summary of the key steps and a complete list of QSS probabilities for all enzyme types are provided in the Supplementary Material. It is important to note that it is necessary to derive the QSS approximation from the original reaction system (Fig.~\ref{Figure1}b) due to the stochastic nature of our model, which requires explicit computation of QSS probabilities. 

In the Supplementary Material, we present pseudocode for the algorithm used to simulate the model after the QSS approximation. We also present the mean-field equations associated with our stochastic model.

\subsection{Baseline and externally induced levels of double-strand breaks} \label{main:methods13}

As mentioned above, in Eq~\eqref{eq_sirt_dsb}, $B$ is a dimensionless parameter representing the number of DSBs per cell. Since our model primarily uses nondimensional quantities (see also reference\supercite{stepanova2025understanding}), we focus on the relative increase in DSBs compared to baseline conditions rather than absolute values.

The natural rate of DSBs in cells, as reported in the literature, ranges from approximately 10 to 50 DSBs per cell per day (or per cell cycle) \supercite{vilenchik2003endogenous,yang2023loss}. Although we do not explicitly use dimensional values, we adopt this range as the baseline level of DSBs when modelling increased genomic instability. We denote the baseline rate as $B^*$ and assume that it represents the spontaneous DNA damage caused by background ionising radiation, cell division, or other environmental factors.

Beyond spontaneous damage, DSBs can also be induced by, for example, ionising radiation during cancer treatment. The baseline number of endogenous DSBs, $B^*$, is roughly equivalent to the DNA damage caused by $1.5-2.0$ Gy (Gray) of ionising radiation \supercite{vilenchik2003endogenous}. In cancer therapy, total administered dosages during treatment are significantly higher, varying from $20$ to $70-100$ Gy or more, depending on cancer type (e.g. see \supercite{okunieff1995radiation,lawrence2010radiation,yahalom2015modern,anderson2021updated}). This corresponds to a 10- to 50- or even 100-fold increase in DSBs relative to baseline levels (i.e. within the range [$10B^*,100B^*]$).

In \supercite{yang2023loss}, non-mutagenic DSBs were introduced in mice (and mouse cells) using the ICE system. Tissues in the ICE mice experienced 10- to 30-fold increases in DSBs compared to untreated wild-type mice (i.e. within [$10B^*,30B^*]$). For the analysis and simulations performed in this study, we view a 20-fold increase in DSBs as moderate genomic instability, a 50-fold increase as high instability, and a maximum of 100-fold increase as extreme genomic instability.

\section*{Data availability}
All relevant data are within the manuscript and its Supplementary Material.

\section*{Code availability}
The detailed description of our model provided in this manuscript and/or in the Supplementary Material (see also pseudocode used for our model simulation outlined in Algorithm 1) allows for the reproducibility of all the presented results. The underlying C++ code for this study is not publicly available but may be made available on reasonable request from the corresponding author.

\section*{Acknowledgements}
D.S. and T.A. thank the CERCA Program/Generalitat de Catalunya for institutional support. D.S. and T.A. have been funded by the grant PID2021-127896OB-I00 funded by MCIN/AEI/10.13039/501100011033 `ERDF A way of making Europe'. The work of D.S. and T.A. has been supported by the Spanish Research Agency (AEI), through the Severo Ochoa and Maria de Maeztu Program for Centres and Units of Excellence in R\&D (CEX2020-001084-M). The funders played no role in study design, analysis and interpretation of modelling results, or the writing of this manuscript. 

\section*{Author contributions}
D.S.: conceptualisation, methodology, model simulations and analysis, writing (original draft, review and editing). H.M.B: conceptualisation, interpretation of modelling results, writing (review and editing), funding acquisition. T.A.: conceptualisation, methodology, interpretation of modelling results, writing (review and editing), funding acquisition. All authors read and approved the final manuscript.

\section*{Competing interests}
All authors declare no financial or non-financial competing interests.

\printbibliography

%\bibliographystyle{apacite}% unsrt apacite
%{\scriptsize{
%\bibliography{reference.bib}
%}}

\end{document}

\title{ \vspace*{0cm} \textbf{Stochastic modelling reveals that chromatin folding buffers epigenetic landscapes against sirtuin depletion during DNA damage -- Supplementary Material}}

\author[1,*]{Daria Stepanova}%
\author[2,3]{Helen M. Byrne}%
\author[4,1,5]{Tom\'{a}s Alarc\'{o}n}%

\affil[1]{Centre de Recerca Matem\`{a}tica, Bellaterra (Barcelona), Spain}%
\affil[2]{Wolfson Centre for Mathematical Biology, Mathematical Institute, University of Oxford, Oxford, UK}%
\affil[3]{Ludwig Institute for Cancer Research, Nuffield Department of Medicine, University of Oxford, Oxford, UK}
\affil[4]{Instituci\'{o} Catalana de Recerca i Estudis Avan\c{c}ats, Barcelona, Spain}%
\affil[5]{Departament de Matem\`{a}tiques, Universitat Aut\`{o}noma de Barcelona, Bellaterra (Barcelona), Spain}%

% \affil[$\dagger$]{These authors contributed equally to this work.}
\affil[*]{\url{dstepanova@crm.cat}}

\date{\today}
\begingroup
\let\center\flushleft
\let\endcenter\endflushleft
\maketitle
\endgroup

%% This is to use Roman numbering for sections and subsections to avoid confusion with the main text
\renewcommand{\thesection}{\Roman{section}} 

\startcontents[sections]
\printcontents[sections]{l}{1}{\setcounter{tocdepth}{2}}

\renewcommand{\thefigure}{S\arabic{figure}}
\setcounter{figure}{0}
\setcounter{equation}{0}

\section{Detailed model description} \label{sec:detailed_model}

\subsection{Notation and full model reactions} \label{SM:sec11}

In this paper, we use the notation detailed in Table~\ref{table_Notation} to describe the dependent variables that appear in our model for epigenetic regulation.

\begin{table}[b!]
\centering
\begin{tabular}{|p{1.5cm} | p{14.0cm}|}
\hline
\textbf{Variable} & \textbf{Description} \\ \hline
$S_{27_i}$ & Unmodified lysine residue 27 of histone H3, H3K27, at $i$th genomic coordinate. \\
$S_{4_i}$ & Unmodified lysine residue 4 of histone H3, H3K4, at $i$th genomic coordinate. \\ \hline
$P_{I_i}$ & H3K27me3 (trimethylation of lysine 27 on histone H3) at $i$th genomic site is a repressive epigenetic mark associated with chromatin silencing. Its accumulation typically leads to chromatin compaction, reducing gene expression in the affected region. \\
$P_{A_i}$ & H3K4me3 (trimethylation of lysine 4 on histone H3) at the $i$th genomic site is an activating epigenetic mark linked to chromatin relaxation. Its accumulation promotes chromatin opening, increasing DNA accessibility and facilitating gene expression. \\
$Q_{{}_i}$ & H3K27ac (acetylation of lysine 27 on histone H3) at the $i$th genomic site is an activating epigenetic mark linked to chromatin relaxation. Its accumulation promotes chromatin opening, increasing DNA accessibility and facilitating gene expression. \\ \hline
$E_J$ & Unbound enzymes of type $J$, with $J \in \uf{Z,M,C,U,K,S}$. The writer enzymes that facilitate deposition of epigenetic marks are $Z$: EZH2 (Enhancer of Zeste 2) of Polycomb Repressor Complex 2, $M$: MLL (Mixed-Lineage Leukemia) of Trithorax family of enzymes, $C$: CBP (CREB-binding protein). The eraser enzymes which mediate the removal of histone modifications are $U$: UTX (Ubiquitously Transcribed Tetratricopeptide Repeat on chromosome X), $K$: KDM (histone lysine demethylase), $S$: NuRD (Nucleosome Remodelling and Deacetylase) and SIRT (sirtuin family). More details on all the enzyme families used in this work can be found in references\supercite{sneppen2019theoretical,stepanova2025understanding}. \\ \hline
$C_{J_i}$ & Complex of enzyme of type $J$ bound to a lysine residue at $i$th genomic site.\\
$C_{SB}$ &  Complex of sirtuin enzyme bound to a site of DSB repair. \\ \hline
\end{tabular}
\caption{\textbf{Notation and description of model variables used in this work.} Where appropriate, the subscript $i$ indicates the index of the genomic locus, $i = 1, \ldots, N$, with $N$ being the total number of genomic loci.}
\label{table_Notation}
\end{table}

The stochastic model comprises $6N+1$ enzymatic reactions, as outlined in Eqs~\eqref{eq_add_marks_27me}-\eqref{eq_sirtuin_B} below, where $N$ is the number of genomic regions. The first $3N$ reactions describe the addition of epigenetic marks to unmodified lysine residues (Eqs~\eqref{eq_add_marks_27me}-\eqref{eq_add_marks_27ac}), while Eqs~\eqref{eq_remove_marks_27me}-\eqref{eq_remove_marks_27ac} describe the removal of these modifications. The reactions in Eqs~\eqref{eq_add_marks_27me}-\eqref{eq_remove_marks_27ac} are identical to those introduced in our previous work \supercite{stepanova2025understanding}. The final reaction (Eq~\eqref{eq_sirtuin_B}) accounts for the sequestration of sirtuins, which typically regulate the acetylation of H3K27 (see Eq~\eqref{eq_remove_marks_27ac}), to sites of double-strand breaks (DSBs). In Eq~\eqref{eq_sirtuin_B}, the dimensionless parameter $B$ represents the extent of DNA damage.

\begin{align} 
\allowdisplaybreaks
\label{eq_add_marks_27me}
& S_{27_i} + E_Z \stacksymbol{\rightleftarrows}{\textcolor{burgundy}{\bar{k}_{1i}}}{\textcolor{burgundy}{\bar{k}_{2i}}} C_{Z_i} \stacksymbol{\rightarrow}{\textcolor{burgundy}{\bar{k}_{3}}}{} P_{I_i} + E_Z, \\ \label{eq_add_marks_4me}
& S_{4_i} + E_M \stacksymbol{\rightleftarrows}{\textcolor{burgundy}{\bar{k}_{4i}}}{\textcolor{burgundy}{\bar{k}_{5i}}} C_{M_i} \stacksymbol{\rightarrow}{\textcolor{burgundy}{\bar{k}_{6}}}{} P_{A_i} + E_M, \\  \label{eq_add_marks_27ac} \displaybreak
& S_{27_i} + E_C \stacksymbol{\rightleftarrows}{\textcolor{burgundy}{\bar{k}_{7i}}}{\textcolor{burgundy}{\bar{k}_{8}}} C_{C_i} \stacksymbol{\rightarrow}{\textcolor{burgundy}{\bar{k}_{9}}}{} Q_{{}_i} + E_C, \\ \label{eq_remove_marks_27me}
& P_{I_i} + E_U \stacksymbol{\rightleftarrows}{\textcolor{burgundy}{\bar{k}_{10}}}{\textcolor{burgundy}{\bar{k}_{11}}} C_{U_i} \stacksymbol{\rightarrow}{\textcolor{burgundy}{\bar{k}_{12}}}{} S_{27_i} + E_U, \\ 
\label{eq_remove_marks_4me}
&  P_{A_i} + E_K \stacksymbol{\rightleftarrows}{\textcolor{burgundy}{\bar{k}_{13}}}{\textcolor{burgundy}{\bar{k}_{14}}} C_{K_i} \stacksymbol{\rightarrow}{\textcolor{burgundy}{\bar{k}_{15}}}{} S_{4_i} + E_K, \\ \label{eq_remove_marks_27ac}
& Q_{{}_i} + E_S \stacksymbol{\rightleftarrows}{\textcolor{burgundy}{\bar{k}_{16}}}{\textcolor{burgundy}{\bar{k}_{17}}} C_{S_i} \stacksymbol{\rightarrow}{\textcolor{burgundy}{\bar{k}_{18}}}{} S_{27_i} + E_S, \\ \label{eq_sirtuin_B}
& B + E_S \stacksymbol{\rightleftarrows}{\textcolor{burgundy}{\bar{k}_{19}}}{\textcolor{burgundy}{\bar{k}_{20}}} C_{SB}.
\end{align}
\noindent In Eqs~\eqref{eq_add_marks_27me}-\eqref{eq_sirtuin_B}, $\bar{k}_r$ for $r \in\uf{1i,2i,3,4i,5i,6,7i,8,\ldots,20}$, $i = 1,\ldots,N$, represent the rate constants associated with the corresponding reactions. We note that $\bar{k}_{19}$ and $\bar{k}_{20}$ are denoted as $\bar{k}_{1}$ and $\bar{k}_{-1}$, respectively, in the main text of the manuscript. The subscript $i$ in certain rate constants indicates that these reaction rates may depend on the genomic coordinate $i$. Specifically, these rate functions incorporate feedback mechanisms, where the addition of new epigenetic modifications is influenced by the presence of existing marks. Their functional forms are given as follows \supercite{stepanova2025understanding}:

\begin{align}
\bar{k}_{1i} & = \bar{k}_1 \of{\bar{k}_{1_0} + \displaystyle \sum_{l=1}^{N} {\overline{w}_{il}} P_{I_l}}, \qquad &&
\bar{k}_{2i}  = \bar{k}_2 \of{\bar{k}_{2_0} + \displaystyle \sum_{l=1}^{N} {\overline{w}_{il}} P_{A_l}}, \nonumber \\
\bar{k}_{4i} & = \bar{k}_4 \of{\bar{k}_{4_0} + \displaystyle \sum_{l=1}^{N} {\overline{w}_{il}} Q_{{}_l}}, &&
\bar{k}_{5i}  = \bar{k}_5 \of{\bar{k}_{5_0} + \displaystyle \sum_{l=1}^{N} {\overline{w}_{il}} P_{I_l}}, \label{eq_kinetic_constants} \\
\bar{k}_{7i} & = \bar{k}_7 \of{\bar{k}_{7_0} + \displaystyle \sum_{l=1}^{N} {\overline{w}_{il}} P_{A_l}}. \nonumber
\end{align}
\noindent In Eqs~\eqref{eq_kinetic_constants}, the timescales of the corresponding reactions are determined by $\of{\bar{k}_{r}}^{-1}$ for $r\in \uf{1,2,4,5,7}$. The terms in parentheses represent the combined effect of baseline enzymatic activity ($\bar{k}_{r_0}$) and the recruited activity influenced by epigenetic modifications in the neighbourhood of the $i$th genomic site, which can attract additional histone-modifying enzymes. The levels of epigenetic marks under the summation signs in Eq~\eqref{eq_kinetic_constants} are weighted by the interaction strength between the target site $i$ and its neighbouring site $l$, denoted as  $\bar{w}_{il}$. Following Stepanova et al.\supercite{stepanova2025understanding}, we assume that the intensity of these interactions is proportional to the contact frequency due to chromatin folding, represented by the matrix $W$, and the levels of H3K27me3 at the relevant sites, i.e. $\bar{w}_{il} = w_{il} P_{I_i} P_{I_l}$. This assumption is supported by experimental evidence indicating that the formation of long-range chromatin interactions, such as loops, is facilitated by Polycomb Repressor Complex 1 (PRC1), which is recruited by H3K27me3 epigenetic marks \supercite{loubiere2019cell}.

We note that the model operates under the assumption of enzyme competition, meaning that enzyme molecules diffuse freely and can bind to any genomic site. A novel feature of the current model is the inclusion of sirtuins, which we assume can exist in three distinct states: \textit{(i)} free or unbound, \textit{(ii)} bound to a genomic site, and \textit{(iii)} sequestered at a DNA repair site. We also assume that the total number of lysine residues at each genomic coordinate remains constant. These conservation laws are expressed as follows:

\begin{align} \label{conservation_law_lysine27}
& S_{27_0}  = S_{27_i} + P_{I_i} + Q_{{}_i} + C_{Z_i} + C_{U_i} + C_{C_i} + C_{S_i}, \\ \label{conservation_law_lysine4}
& S_{4_0}  = S_{4_i} + P_{A_i} + C_{M_i} + C_{K_i},\\ \label{conservation_law_enzyme}
& E_J + \displaystyle \sum_{l=1}^N C_{J_l} = E_{J_0}, \qquad J \in \uf{M,Z,C,K,U}, \\
& E_S + C_{SB} + \displaystyle \sum_{l=1}^N C_{S_l} = E_{S_0}. \label{conservation_law_sirtuin}
\end{align}
\noindent Here, the positive constants $S_{27_0}$ and $S_{4_0}$ represent the total number of H3K27 and H3K4 residues at genomic site $i = 1, \ldots, N$. The total amount of enzyme of type $J$ is denoted by $E_{J_0}$, where $J \in \uf{M, Z, C, K, U, S}$.

In summary, our stochastic model is defined by Eqs~\eqref{eq_add_marks_27me}-\eqref{conservation_law_sirtuin}, along with a suitable selection of parameter values and initial conditions for all variables.

\subsection{Quasi-steady state approximation and model simulation} \label{SM:sec12}

The model presented in Section~\ref{SM:sec11} can be simplified by assuming that histone-modifying enzymes are significantly less abundant than their binding sites, which are represented by lysine residues. Specifically, we define the characteristic scalings of lysine residues as $\Omega_S$ and those of histone-modifying enzymes as $\Omega_E$, and impose the relationship $\Omega_E \ll \Omega_S$ reflecting the low availability of the enzyme relative to the substrate. This assumption is supported by experimental studies \supercite{stafford2018multiple} and theoretical modelling by Owen et al., which illustrates how a limited supply of reader-writer enzymes helps maintain epigenetic memory and sustain stable epigenetic patterns \supercite{owen2023design}. Under this scaling, enzyme-substrate complexes form on a timescale which is much shorter than that on which epigenetic modifications occur. We use the Briggs-Haldane approximation \supercite{briggs1925note}, which justifies a separation of timescales: enzyme binding and unbinding operate on a fast timescale, while the modification of epigenetic marks progresses more slowly. Under these assumptions, the model can be simplified to describe only the evolution of epigenetic modifications, with enzyme complexes rapidly equilibrating to their quasi-steady-state (QSS) distributions. The reactions in the reduced model are formulated as follows:

\begin{align} \label{marks_reduced}
& S_{27_i} \stacksymbolL{\rightleftarrows}{\textcolor{burgundy}{{\small{\kappa_3 c_{Z_i}}}}}{\textcolor{burgundy}{{\small{\kappa_{12} c_{U_i}}}}} P_{I_i} , \qquad \qquad
S_{4_i}  \stacksymbolL{\rightleftarrows}{\textcolor{burgundy}{{\small{\kappa_{6} c_{M_i}}}}}{\textcolor{burgundy}{{\small{\kappa_{15} c_{K_i}}}}} P_{A_i}, \qquad \qquad
S_{27_i} \stacksymbolL{\rightleftarrows}{\textcolor{burgundy}{{\small{\kappa_{9} c_{C_i}}}}}{\textcolor{burgundy}{{\small{\kappa_{18} c_{S_i}}}}} Q_{{}_i}.
\end{align}

\noindent These reactions can be simulated using the Stochastic Simulation Algorithm (e.g. \supercite{gillespie1976general}), where the number of enzyme-substrate complexes is dynamically sampled from their corresponding quasi-steady-state (QSS) distributions at each simulation step. In Eq~\eqref{marks_reduced}, these complexes are normalised by their characteristic scaling factor, $\Omega_E$, such that $c_{J_i} = C_{J_i} \Omega_E^{-1}$. 

The rescaled kinetic constants in Eq~\eqref{marks_reduced} and the QSS distributions shown below are coupled to the kinetic parameters of the full model through the following relationships:  
\begin{itemize} \setlength{\itemsep}{5pt}
\item $\kappa_r = \bar{k}_r \bar{k}_7^{-1} \Omega_S^{-2}$ for $r \in \uf{3,6,8,9,11,12,14,15,17,18,19,20}$, 
\item $\kappa_r = \bar{k}_r \bar{k}_7^{-1} \Omega_S^{-1}$ for $r \in \uf{2,5,10,13,16}$, 
\item $\kappa_{r_0} = \bar{k}_{r_0} \Omega_S^{-1}$ for $r \in \uf{1,2,4,5,7}$, 
\item $\kappa_r = \bar{k}_r \bar{k}_7^{-1}$ for $r \in \uf{1,4}$.
\end{itemize}

\noindent These expressions are derived as part of the QSS approximation; see reference\supercite{stepanova2025understanding} for a detailed derivation. Acknowledging for the sequestration of sirtuins at DSB repair sites affects the QSS distribution specific to this enzyme class. However, since the distributions of all the enzymes are needed for model simulations, we state them here for completeness.

As in reference\supercite{stepanova2025understanding}, the QSS distributions for free enzymes and enzyme-substrate complexes (where enzymes exist in either a free state or bound to a histone tail) follow multinomial distributions. As such, the probability density function for enzymes of type $J$, where $J \in \uf{M, Z, C, K, U}$, is given by the following expression:

\begin{align} \label{dist_multinomial_enzymeJ}
\phi_{J} (E_J, C_{J_1},\ldots,C_{J_N} \mid \mathbf{p_I},\mathbf{p_A},\mathbf{q}) = \frac{\of{E_{J_0}}!}{\of{E_J}! ~ \displaystyle \prod_{l = 1}^N \of{C_{J_l}}!} ~ \of{p_{E_J}}^{E_J} ~ \displaystyle \prod_{l = 1}^N \of{p_{C_{J_l}}}^{C_{J_l}}.
\end{align}
In Eq~\eqref{dist_multinomial_enzymeJ}, we denote by $\mathbf{p_I} = \uf{p_{I_1}, \ldots, p_{I_N}}$, $\mathbf{p_A} = \uf{p_{A_1}, \ldots, p_{A_N}}$ and $\mathbf{q} = \uf{q_{{}_1}, \ldots, q_{{}_N}}$ levels of the three epigenetic modifications along the chromatin, with the normalised concentrations of marks given by $p_{I_i} = P_{I_i} \Omega_S^{-1}$, $p_{A_i} = P_{A_i} \Omega_S^{-1}$ and $q_{I_i} = Q_{{}_i} \Omega_S^{-1}$. We denote by $p_{E_J}$ the QSS probability of enzyme $J$ existing in a free state, and by $p_{C_{J_l}}$ the probability that it is bound to genomic site $l$. The functional forms for each enzyme type can be written as follows:

\begin{subequations} \label{dist_multinomial_enzyme}
\begin{align}
\allowdisplaybreaks
&p_{E_Z} = \frac{1}{1+\displaystyle \sum_j k_{1_j}(\mathbf{p_I},\mathbf{p_A}) (s_{27_0} - p_{I_j} - q_{{}_j})},  \qquad && p_{C_{Z_i}} = \frac{k_{1_i}(\mathbf{p_I},\mathbf{p_A})(s_{27_0} - p_{I_i} - q_{{}_i})}{1+\displaystyle \sum_j k_{1_j}(\mathbf{p_I},\mathbf{p_A})(s_{27_0} - p_{I_j} - q_{{}_j})},   \\[10pt]   
& p_{E_M} = \frac{1}{1+\displaystyle \sum_j k_{4_j}(\mathbf{p_I},\mathbf{q}) (s_{4_0} - p_{A_j})}, && p_{C_{M_i}} = \frac{k_{4_i}(\mathbf{p_I},\mathbf{q})(s_{4_0} - p_{A_i})}{1+\displaystyle \sum_j k_{4_j}(\mathbf{p_I},\mathbf{q}) (s_{4_0} - p_{A_j})},  \\[10pt] 
&p_{E_C} = \frac{1}{1+\displaystyle \sum_j k_{7_j}(\mathbf{p_I},\mathbf{p_A}) (s_{27_0} - p_{I_j} - q_{{}_j})}, && p_{C_{C_i}} = \frac{k_{7_i}(\mathbf{p_I},\mathbf{p_A})(s_{27_0} - p_{I_i} - q_{{}_i})}{1+\displaystyle \sum_j k_{7_j}(\mathbf{p_I},\mathbf{p_A}) (s_{27_0} - p_{I_j} - q_{{}_j})},   \\[10pt] %\displaybreak
&p_{E_U} = \frac{1}{1+ \alpha_{10} \displaystyle \sum_j  p_{I_j}}, && p_{C_{U_i}} = \frac{ \alpha_{10} \skp p_{I_i} }{1+ \alpha_{10}  \displaystyle \sum_j p_{I_j} }, \\[10pt]
& p_{E_K} = \frac{1}{1+ \alpha_{13} \displaystyle \sum_j  p_{A_j}}, && p_{C_{K_i}} = \frac{ \alpha_{13} \skp p_{A_i} }{1+\alpha_{13} \displaystyle \sum_j p_{A_j} }.
\end{align}
\end{subequations}
\noindent In Eqs~\eqref{dist_multinomial_enzyme}, we exploit the normalised conservation laws (Eqs~\eqref{conservation_law_lysine27}-\eqref{conservation_law_lysine4}) which yield $s_{27_i} = S_{27_i} \Omega_S^{-1} = s_{27_0} - p_{I_i} - q_{{}_i} + \mathcal{O}(\epsilon) \approx s_{27_0} - p_{I_i} - q_{{}_i}$ and $s_{4_i} = S_{4_i} \Omega_S^{-1} = s_{4_0} - p_{A_i} + \mathcal{O}(\epsilon) \approx s_{4_0} - p_{A_i}$, for any genomic site $i = 1,\ldots, N$. The summations are over all genomic coordinates, $j = 1,\ldots,N$. Finally, we also introduced the following notation:

\begin{align}
\allowdisplaybreaks
& k_{1_i}(\mathbf{p_I},\mathbf{p_A}) = \frac{\alpha_{I_0} + \kappa_1 \displaystyle \sum_l \overline{w}_{il} \skpp p_{I_l}}{\overline{\kappa}_3 + \kappa_2 \displaystyle \sum_l \overline{w}_{il} \skpp p_{A_l}}, \hspace{5pt} && \overline{\kappa}_3 = \kappa_3 + \kappa_2 \kappa_{2_0}, \hspace{3pt} &&& \alpha_{I_0} = \kappa_1 \kappa_{1_0}, \hspace{22pt}  \nonumber\\
& k_{4_i}(\mathbf{p_I},\mathbf{q}) = \frac{\alpha_{A_0} + \kappa_4 \displaystyle \sum_l \overline{w}_{il} \skpp q_{{}_l}}{\overline{\kappa}_6 + \kappa_5 \displaystyle \sum_l \overline{w}_{il} \skpp p_{I_l} },  && \overline{\kappa}_6 = \kappa_6 + \kappa_5 \kappa_{5_0},  &&& \alpha_{A_0} = \kappa_4 \kappa_{4_0},  \hspace{22pt}  \label{eq_meanfield_constants}  \\ \displaybreak
& k_{7_i}(\mathbf{p_I},\mathbf{p_A}) = \frac{1}{\alpha_8}\of{\alpha_{Q_0} +  \displaystyle \sum_l \overline{w}_{il} \skpp p_{A_l}},  && \alpha_8 = \kappa_8 + \kappa_9,  &&& \alpha_{Q_0} = \kappa_{7_0}, \hspace{33pt}   \nonumber \\
& \alpha_{10} = \frac{\kappa_{10}}{\kappa_{11}+\kappa_{12}}, &&  \alpha_{13} = \frac{\kappa_{13}}{\kappa_{14}+\kappa_{15}}. \nonumber
\end{align}
\noindent The effective kinetic constants, $k_{r_i}$, explicitly incorporate their dependence on the epigenetic landscape variables $\mathbf{p_I}$, $\mathbf{p_A}$ and $\mathbf{q}$. In Eq~\eqref{eq_meanfield_constants}, the interaction strength between genomic sites $i$ and $l$ is represented as $\overline{w}_{il} = w_{il} p_{I_i} p_{I_l}$, where $w_{il}$ is the model input that defines chromatin folding.

The sirtuin family of enzymes exists in three states: free, bound to a genomic site (like other enzymes), or sequestered in a complex at a DNA repair site. By following the approach used to derive the QSS approximation (see reference\supercite{stepanova2025understanding}), it is possible to show that the QSS distribution of sirtuins also follows a multinomial distribution, with a probability density function given by

\begin{align} \label{dist_multinomial_enzymeS}
\phi_{S} (E_S, C_{SB}, C_{S_1},\ldots,C_{S_N} \mid \mathbf{p_I},\mathbf{p_A},\mathbf{q}) = \frac{\of{E_{S_0}}!}{\of{E_S}! ~ \of{C_{SB}}! ~ \displaystyle \prod_{l = 1}^N \of{C_{S_l}}!} ~ \of{p_{E_S}}^{E_S} ~ \of{p_{C_{SB}}}^{C_{SB}} ~ \displaystyle \prod_{l = 1}^N \of{p_{C_{S_l}}}^{C_{S_l}},
\end{align}
\noindent where the QSS probabilities are defined as follows:

\begin{align} \label{eq_QSS_prob_sirtuin}
p_{E_S} &= \frac{1}{1+\alpha_{19} B + \alpha_{16} \displaystyle \sum_l  q_{{}_l}}, \qquad p_{E_{SB}} = \frac{\alpha_{19} B}{1+\alpha_{19} B + \alpha_{16} \displaystyle \sum_l  q_{{}_l}}, \qquad p_{C_{S_i}} = \frac{\alpha_{16} q_{{}_i} }{1+ \alpha_{19} B + \alpha_{16} \displaystyle \sum_l q_{{}_l} }.
\end{align}
\noindent In Eq~\eqref{eq_QSS_prob_sirtuin}, $\alpha_{16} = \kappa_{16}\of{\kappa_{17}+\kappa_{18}}^{-1}$ and $\alpha_{19} = \kappa_{19} \kappa_{20}^{-1}$.

These QSS distributions complete the formulation of our reduced model, following the timescale separation. To ensure that the system's behaviour remains invariant to the number of genomic sites into which the chromatin region is divided, we must apply the scaling approach outlined in reference\supercite{stepanova2025understanding}. Specifically, in the QSS probabilities given by Eqs~\eqref{dist_multinomial_enzyme} and \eqref{eq_QSS_prob_sirtuin}, all terms of the form $p_{I_i}$, $p_{A_i}$, $q_{{}_i}$, $(s_{4_0} - p_{A_i})$, or $(s_{27_0} - p_{I_i} - q_{{}_i})$ must be scaled by a factor of $2/N$ (if initially the chromatin region is divided into two subregions).

In Algorithm~\ref{Alg1}, we present pseudocode outlining the main steps for its stochastic model simulation. The algorithm follows the standard Stochastic Simulation Algorithm (SSA) \supercite{gillespie1976general}, with the key distinction that at each timestep, enzymatic complexes are drawn from their QSS distributions.

\begin{algorithm}[t]
\caption{Stochastic simulation algorithm for the epigenetic regulation model}
\label{alg:Gillespie}
\begin{algorithmic}[1]
\State \textbf{Model input:} Chromatin contact map, $W$, parameter values and the characteristic levels of lysine residues and enzymes in the system ($\Omega_S$ and $\Omega_E$). 
\State \textbf{Initialise:} Set initial conditions for epigenetic marks $\mathbf{X} = \of{\mathbf{p_I},\mathbf{p_A},\mathbf{q}}$, the dimension of $\mathbf{X}$ is $3N$, where $N$ is the total number of genomic sites.
\State Set total simulation time $T_{\text{max}}$, initialise $t = 0$.
\While{$t < T_{\text{max}}$}
    \State Draw the vectors of enzymatic complexes, $\of{C_{J_1},\ldots,C_{J_N}}$, of type $J \in \uf{M,Z,C,K,U}$ from the corresponding 
    
multinomial distributions (Eqs~\eqref{dist_multinomial_enzymeJ}-\eqref{dist_multinomial_enzyme}). Enzymatic complexes of the sirtuin family are sampled from the

multinomial distribution given by Eqs~\eqref{dist_multinomial_enzymeS}-\eqref{eq_QSS_prob_sirtuin}.
    \State Calculate reaction propensities $a_{R_i}(\mathbf{X})$ for six reactions ($R = 1,\dots,6$) which can occur in $i = 1,\ldots,N$ genomic 
    
sites. \Comment{see Eqs~\eqref{eq_propensity_deposition}-\eqref{eq_propensity_removal}}
    \State Compute total propensity: $a_{\text{tot}} = \sum_{R=1}^{6} \sum_{i=1}^{N} a_{R_i}(\mathbf{X})$.
    \State Draw two random numbers $r_1, r_2 \sim \text{Uniform}(0,1)$.
    \State Compute the time step for the next reaction to occur: $\Delta t = \frac{1}{a_{\text{tot}}} \ln(1/r_1)$.
    \State Select reaction indexes $R^*$ and $i^*$ such that:
    \[
    \sum_{R=1}^{R^*} \sum_{i=1}^{i^*-1} a_{R_i} < r_2 a_{\text{tot}} \leq \sum_{R=1}^{R^*} \sum_{i=1}^{i^*} a_{R_i}
    \]
    \State Update system state: $\mathbf{X} \gets \mathbf{X} + \boldsymbol{\nu}$, where $ \boldsymbol{\nu}$ is the stoichiometric vector corresponding to the selected reaction. 
    
If $R^* \in \uf{1,2,3}$ (reactions for mark deposition), then the non-zero vector component $\mathbf{\nu}_{j^*} = +1$, 

where $j^* = R^* \cdot i^*$. 

If $R^* \in \uf{4,5,6}$ (reactions for mark removal), then the non-zero vector component $\mathbf{\nu}_{j^*} = -1$, 

where $j^* = (R^*- 3) \cdot i^*$. 

\Comment{see a more detailed description within the text}
    \State Update time: $t \gets t + \Delta t$.
\EndWhile
\State \textbf{Output:} Evolution of the epigenetic landscape, $\mathbf{X} = \of{\mathbf{p_I},\mathbf{p_A},\mathbf{q}}$, over time.\end{algorithmic}
\label{Alg1}
\end{algorithm}

The reaction propensities corresponding to Eq~\eqref{marks_reduced} which we use in Algorithm~\ref{Alg1} are defined as follows:

\begin{align} \label{eq_propensity_deposition}
\text{\textbf{Mark deposition:}} \quad & \alpha_{1_i} = \Omega_S \kappa_3 C_{Z_i} \Omega_E^{-1}, \quad && \alpha_{2_i} = \Omega_S \kappa_6 C_{M_i} \Omega_E^{-1}, \quad && \alpha_{3_i} = \Omega_S \kappa_9 C_{C_i} \Omega_E^{-1} ;& \\
\text{\textbf{Mark removal:}} \quad & \alpha_{4_i} = \Omega_S \kappa_{12} C_{U_i} \Omega_E^{-1},   && \alpha_{5_i} = \Omega_S \kappa_{15} C_{K_i} \Omega_E^{-1},  && \alpha_{6_i} = \Omega_S \kappa_{18} C_{S_i} \Omega_E^{-1}. \label{eq_propensity_removal}
\end{align}

We also define the stoichiometric vectors, $\boldsymbol{\nu}$, to represent changes in the system’s configuration, specifically the variation in the number of epigenetic marks along the chromatin region. The vector is given by $\boldsymbol{\nu} = (\nu_1, \ldots, \nu_{3N})$, where each entry corresponds to a change in a specific epigenetic mark at a particular genomic coordinate. Its first $N$ entries correspond to changes in H3K27me3 ($\mathbf{p_I}$). Entries $(N+1)$ to $2N$ correspond to changes in H3K4me3 ($\mathbf{p_A}$). The final $N$ components represent changes in H3K27ac ($\mathbf{q}$) landscape. Since each reaction modifies only one epigenetic mark at a single genomic coordinate, the stoichiometric vector contains a single nonzero entry, with all other components set to zero. We denote this nonzero component by $\nu_{j^*} = \pm 1$, where $j^*$ specifies the affected mark and its corresponding genomic site. In Algorithm~\ref{Alg1}, we explain how to calculate $j^*$ for the particular reactions of our model.

For all numerical simulations of our stochastic model presented in this study, we use Eq~\eqref{marks_reduced} and Algorithm~\ref{Alg1}.

\subsection{Mean-field limit} \label{SM:sec13}

To investigate how our model responds to changes in parameter values, we derive the mean-field equations associated with the reduced stochastic system. While this analysis is feasible only for simplified scenarios (see also Section~\ref{SM:sec14}), it provides valuable insight into the system's behaviour. Since the derivation of the mean-field equations from Eqs~\eqref{marks_reduced}-\eqref{eq_QSS_prob_sirtuin} follows the same steps as in our previous work \supercite{stepanova2025understanding}, we focus here only on the final system of ordinary differential equations:

\begin{align}
\textbf{H3K27me3:}~~ \quad &\dv{p_{I_i}}{\tau} =  \frac{\kappa_3 e_{Z_0} k_{1_i}(\mathbf{p_I},\mathbf{p_A})\of{s_{27_0} - p_{I_i} - q_{{}_i}}}{1+\displaystyle \sum_j  k_{1_j}(\mathbf{p_I},\mathbf{p_A}) \of{s_{27_0} - p_{I_j} - q_{{}_j}}}  - \frac{\kappa_{12} e_{U_0} \alpha_{10} ~p_{I_i}}{1+ \alpha_{10} \displaystyle \sum_j p_{I_j}}, \label{eq_marks_meanfield_final1} \\
\textbf{H3K4me3:} ~~~ \quad &\dv{p_{A_i}}{\tau} =  \frac{\kappa_6 e_{M_0} k_{4_i}(\mathbf{p_I},\mathbf{q}) \of{s_{4_0} - p_{A_i}}}{1+\displaystyle \sum_j  k_{4_j}(\mathbf{p_I},\mathbf{q}) \of{s_{4_0} - p_{A_j}}}  - \frac{\kappa_{15} e_{K_0} \alpha_{13} ~p_{A_i}}{1+ \alpha_{13} \displaystyle \sum_j p_{A_j}},  \label{eq_marks_meanfield_final2} \\
\textbf{H3K27ac:} ~~~ \quad&\dv{q_{{}_i}}{\tau} ~ =  \frac{\kappa_9 e_{C_0} k_{7_i}(\mathbf{p_I},\mathbf{p_A})\of{s_{27_0} - p_{I_i} - q_{{}_i}}}{1+\displaystyle \sum_j  k_{7_j}(\mathbf{p_I},\mathbf{p_A}) \of{s_{27_0} - p_{I_j} - q_{{}_j}}}  - \frac{\kappa_{18} e_{S_0} \alpha_{16} ~q_{{}_i}}{1+ \alpha_{19} B + \alpha_{16} \displaystyle \sum_j q_{{}_j}}. \label{eq_marks_meanfield_final3}
\end{align}
\noindent In Eqs~\eqref{eq_marks_meanfield_final1}-\eqref{eq_marks_meanfield_final3}, the effective kinetic constants, $k_{r_i}$ for $r \in \uf{1,4,7}$, are as defined in Eq~\eqref{eq_meanfield_constants} and the indexes $i$ and $j$ vary over all the genomic coordinates $1, \ldots, N$.

The last term in Eq~\eqref{eq_marks_meanfield_final3} represents the removal of acetylation groups from H3 histone tails. Specifically, the removal of H3K27ac marks is attenuated due to the redistribution of sirtuins from histones to DNA repair sites (represented by a dimensionless quantity, $B$).

Eqs~\eqref{eq_marks_meanfield_final1}-\eqref{eq_marks_meanfield_final3} must be complemented by appropriate initial conditions for epigenetic modifications at each genomic region.

\subsection{Two-region mean-field equations} \label{SM:sec14}

One simple scenario considered in this study involves a mean-field system of equations where the chromatin domain is divided into two distinct regions. The first region contains only diagonal connections, while in the second region, all genomic sites are uniformly interconnected. This interconnected subdomain serves as an approximation of a small topologically associated domain (TAD), a fundamental chromatin conformation \supercite{szabo2019principles}. Under these assumptions, the contact map, $W$, takes the following form:

\begin{itemize} \setlength{\itemsep}{5pt}
\item For $i = 1,\ldots,R$, where $R$ is the size of region 1, $w_{ii} = w_{diag}$ and $w_{ij} = 0$ for all $i\ne j\in \uf{1,\ldots,R}$. 
\item For $i = R+1,\ldots,N$,  $w_{ij} = w_{TAD}$ for all $i,j \in \uf{R+1,\ldots,N}$. The size of the TAD region is, thus, $\bar{R} = (N-R)$.
\item The rest of the values of $W$ are set to zero: $w_{ij} = 0$ and $w_{ji} = 0$ for $i \in \uf{1,\ldots,R}$ and $j \in \uf{R+1,\ldots,N}$.
\end{itemize}
\noindent Eq~\eqref{Wmatrix_tworegions} represents this contact map in matrix form. Additionally, a visual representation of this chromatin architecture is provided in Fig.~\ref{Figure2}a of the main text.

\begin{align} \label{Wmatrix_tworegions}
W =
\begin{pmatrix}
    w_{\text{diag}} & 0 & \cdots & 0 & \vline & 0 & \cdots & 0 \\
    0 & w_{\text{diag}} & \cdots & 0 & \vline & 0 & \cdots & 0 \\
    \vdots & \vdots & \ddots & \vdots & \vline & \vdots & \ddots & \vdots \\
    0 & 0 & \cdots & w_{\text{diag}} & \vline & 0 & \cdots & 0 \\
    \hline
    0 & 0 & \cdots & 0 & \vline & w_{\text{TAD}} & \cdots & w_{\text{TAD}} \\
    \vdots & \vdots & \ddots & \vdots & \vline & \vdots & \ddots & \vdots \\
    0 & 0 & \cdots & 0 & \vline & w_{\text{TAD}} & \cdots & w_{\text{TAD}}
\end{pmatrix}
\end{align}

With the contact map $W$ defined by Eq~\eqref{Wmatrix_tworegions}, the mean-field ODE system (Eqs~\eqref{eq_marks_meanfield_final1}-\eqref{eq_marks_meanfield_final3}) simplifies significantly. Due to the uniform connectivity between genomic sites within the same region, the right-hand side of the equations is identical for all sites within a given region. If we further assume that the initial conditions for the epigenetic marks are uniform within each region (though they may differ between regions), the system reduces to six equations describing the evolution of the three epigenetic marks at two representative genomic sites: one for region 1 and another for region 2. Specifically, we define the representative epigenetic mark levels for the diagonal region as $\widetilde{p}_{I_1} = p_{I_i}$, $\widetilde{p}_{A_1} = p_{A_i}$, and $\widetilde{q}_{{}_1} = q_{{}_i}$ for $i = 1, \ldots, R$. Similarly, for the TAD-like region, we introduce $\widetilde{p}_{I_2} = p_{I_i}$, $\widetilde{p}_{A_2} = p_{A_i}$, and $\widetilde{q}_{{}_2} = q_{{}_i}$ for $i = R+1, \ldots, N$. The derivation of these equations, including the redistribution of sirtuins in the extended model, follows an approach similar to that outlined in reference\supercite{stepanova2025understanding}. Thus, we present only the final set of equations, which read:

\begin{align}
 \label{eq_two_subregions1} 
\dv{\widetilde{p}_{I_{i_s}}}{\tau} &=  \frac{\kappa_3 e_{Z_0} \widetilde{k}_{1_{i_s}} \skpp \of{s_{27_0} - \widetilde{p}_{I_{i_s}} - \widetilde{q}_{{}_{i_s}}}}{ {\color{dblue}{N/2}}+ {\color{dblue}{R}} \skp \widetilde{k}_{1_1} \skpp \of{s_{27_0} - \widetilde{p}_{I_1} - \widetilde{q}_{{}_1}} + {\color{dblue}{\bar{R}}} \skp \widetilde{k}_{1_2} \skpp \of{s_{27_0} - \widetilde{p}_{I_2} - \widetilde{q}_{{}_2}}}  \\ 
& -\frac{\kappa_{12} e_{U_0} \alpha_{10} ~ \widetilde{p}_{I_{i_s}}}{ {\color{dblue}{N/2}} + \alpha_{10} \of{ {\color{dblue}{R}} \skpp \widetilde{p}_{I_1} + {\color{dblue}{\bar{R}}} \skpp \widetilde{p}_{I_2}}}, \nonumber \\
\label{eq_two_subregions2} 
\dv{\widetilde{p}_{A_{i_s}}}{\tau} &=  \frac{\kappa_6 e_{M_0} \widetilde{k}_{4_{i_s}} \of{s_{4_0} - \widetilde{p}_{A_{i_s}}}}{ {\color{dblue}{N/2}} + {\color{dblue}{R}} \skp \widetilde{k}_{4_1} \skpp \of{s_{4_0} - \widetilde{p}_{A_1}} + {\color{dblue}{\bar{R}}} \skp \widetilde{k}_{4_2} \skpp \of{s_{4_0} - \widetilde{p}_{A_2}} } \\
& - \frac{\kappa_{15} e_{K_0} \alpha_{13} \skpp \widetilde{p}_{A_{i_s}}}{ {\color{dblue}{N/2}} + \alpha_{13}\of{ {\color{dblue}{R}} \skp  \widetilde{p}_{A_1} +  {\color{dblue}{\bar{R}}} \skp \widetilde{p}_{A_2} }},  \nonumber \\
 \label{eq_two_subregions3} 
\dv{\widetilde{q}_{{}_{i_s}}}{\tau} &=  \frac{\kappa_9 e_{C_0} \widetilde{k}_{7_{i_s}} \skpp \of{s_{27_0} - \widetilde{p}_{I_{i_s}} - \widetilde{q}_{{}_i}}}{ {\color{dblue}{N/2}} + {\color{dblue}{R}} \skp \widetilde{k}_{7_1} \skpp \of{s_{27_0} - \widetilde{p}_{I_1} - \widetilde{q}_{{}_1}} + {\color{dblue}{\bar{R}}} \skp \widetilde{k}_{7_2} \skpp \of{s_{27_0} - \widetilde{p}_{I_2} - \widetilde{q}_{{}_2}}}  \\
& - \frac{\kappa_{18} e_{S_0} \alpha_{16} \skp \widetilde{q}_{{}_{i_s}}}{{\color{dblue}{N/2}}(1+ \alpha_{19}B)  + \alpha_{16} \of{ {\color{dblue}{R}} \skp  \widetilde{q}_{{}_1} + {\color{dblue}{\bar{R}}} \skp \widetilde{q}_{{}_2}}}. \nonumber
\end{align}
\noindent In this system, $i_s = 1,2$ denotes the region index and the modified effective kinetic rates,  $\widetilde{k}_{r_i}$, are obtained from Eq~\eqref{eq_meanfield_constants} by substituting $W$ from Eq~\eqref{Wmatrix_tworegions}:

\begin{align}
& \widetilde{k}_{1_1} = \frac{\alpha_{I_0} + \kappa_1 w_{diag} ~\of{\widetilde{p}_{I_1}}^3}{\overline{\kappa}_3 + \kappa_2 w_{diag} ~\of{\widetilde{p}_{I_1}}^2 \widetilde{p}_{A_1}}, \hspace{5pt} && k_{1_2} = \frac{\alpha_{I_0} + \kappa_1 {\color{dblue}{\bar{R}}} \skpp w_{TAD} ~\of{\widetilde{p}_{I_2}}^3}{\overline{\kappa}_3 + \kappa_2 {\color{dblue}{\bar{R}}} \skpp w_{TAD} ~\of{\widetilde{p}_{I_2}}^2 \widetilde{p}_{A_2}}, \nonumber \\
& \widetilde{k}_{4_1} = \frac{\alpha_{A_0} + \kappa_4 w_{diag} ~\of{\widetilde{p}_{I_1}}^2 \widetilde{q}_{{}_1}}{\overline{\kappa}_6 + \kappa_5 w_{diag} ~\of{\widetilde{p}_{I_1}}^3},  && \widetilde{k}_{4_2} = \frac{\alpha_{A_0} + \kappa_4 {\color{dblue}{\bar{R}}} \skpp w_{TAD} ~\of{\widetilde{p}_{I_2}}^2 \widetilde{q}_{{}_2}}{\overline{\kappa}_6 + \kappa_5 {\color{dblue}{\bar{R}}} \skpp w_{TAD} ~\of{\widetilde{p}_{I_2}}^3},  \label{eq_rates_subregions_tildek} \\
& \widetilde{k}_{7_1}= \frac{1}{\alpha_8}\of{\alpha_{Q_0} + w_{diag} ~\of{\widetilde{p}_{I_1}}^2 \widetilde{p}_{A_1}},  && \widetilde{k}_{7_2}= \frac{1}{\alpha_8}\of{\alpha_{Q_0} + {\color{dblue}{\bar{R}}} \skpp w_{TAD} ~\of{\widetilde{p}_{I_2}}^2 \widetilde{p}_{A_2}} .  \nonumber
\end{align}

The prefactors highlighted in blue in Eqs~\eqref{eq_two_subregions1}-\eqref{eq_rates_subregions_tildek} arise from two factors: the scaling approach ($N/2$), as detailed in Section~\ref{SM:sec12}, and the summation terms involving the contact matrix $W$ from Eq~\eqref{Wmatrix_tworegions} ($R$ and $\bar{R}$).

\printbibliography

\clearpage
\section{Supplementary figures and table}

\renewcommand{\thetable}{\arabic{table}}
\renewcommand{\tablename}{Supplementary Table}
\setcounter{table}{0}

\renewcommand{\thefigure}{\arabic{figure}}
\renewcommand{\figurename}{Supplementary Figure}
\setcounter{figure}{0}

\begin{figure}[!h]
\begin{adjustwidth}{-0.5cm}{-0.5cm}
\begin{center}
\includegraphics[width=19cm]{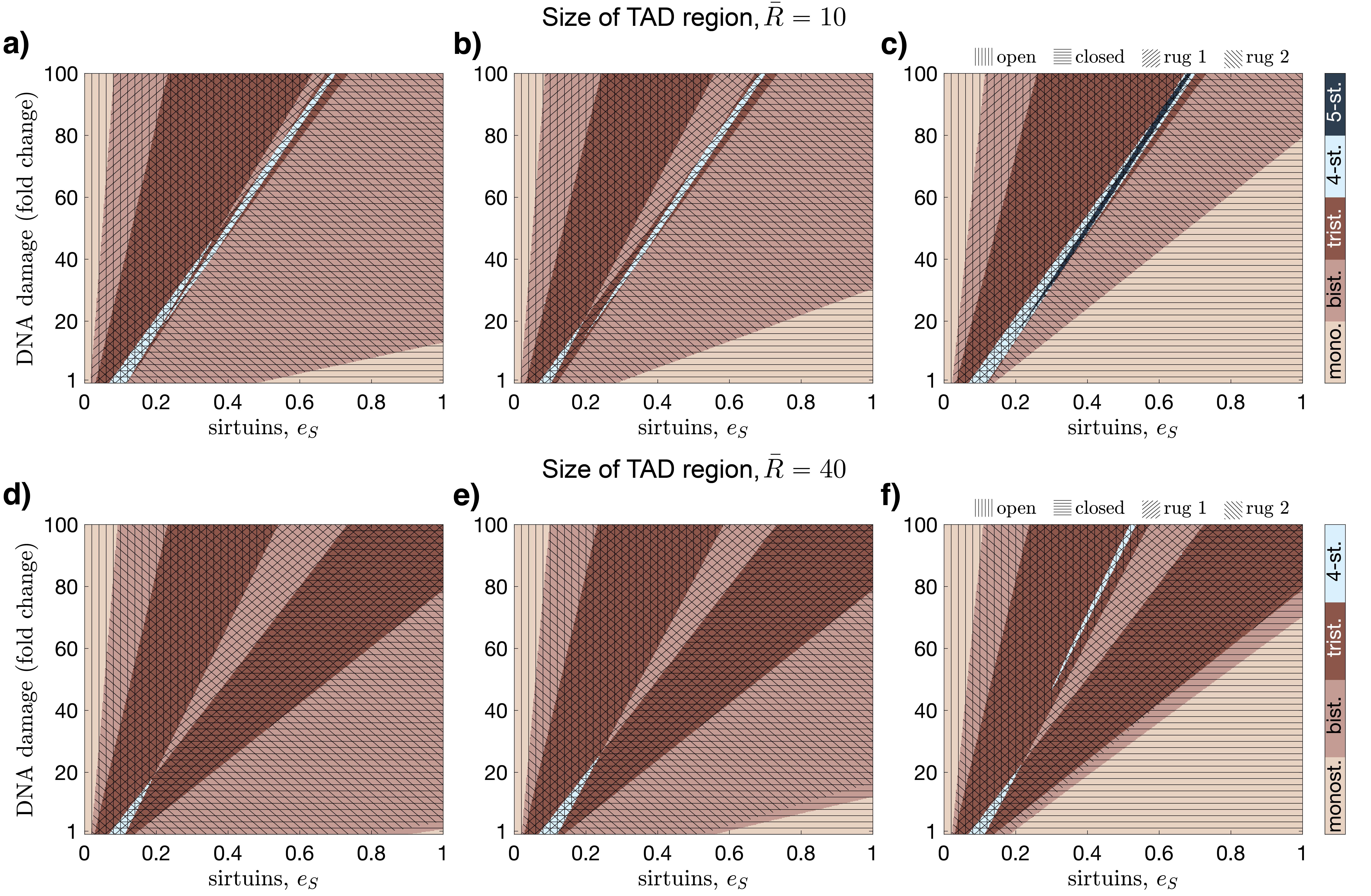}
\caption{\textbf{Phase diagrams of the two-region chromatin system showing how the number and type of stable steady states vary with sirtuin concentration ($e_S$) and increasing DNA damage levels (relative to the baseline $B = B^*$).} The total chromatin domain consists of $N = 50$ genomic sites, with the TAD region occupying \textbf{a-c)} $\bar{R} = 10$ sites and \textbf{d-f)} $\bar{R} = 40$ (results for $\bar{R} = 25$ are presented in Fig.~\ref{Figure3}a-c of the main text). The contact strength in the diagonal region is held constant at $w_{diag} = 100$, while TAD connectivity decreases across columns: \textbf{a, d)} $w_{TAD} = 100$, \textbf{b, e)} $w_{TAD} = 50$, and \textbf{c, f)} $w_{TAD} = 10$. Colour bars indicate the number of coexisting stable steady states; overlaid hatch patterns specify the nature of these states: vertical for uniformly open, horizontal for uniformly closed, diagonal ($\nearrow$) for rugged 1, and diagonal ($\searrow$) for rugged 2. States not matching these four classifications are left unhatched. The corresponding mean-field equations are detailed in Section~\ref{SM:sec14}; all other parameters are listed in Supplementary Table~\ref{SuppTable1}.}
\label{SuppFig1}
\end{center}
\end{adjustwidth}
\end{figure}

\begin{figure}[!h]
\begin{center}
\includegraphics[width=14.5cm]{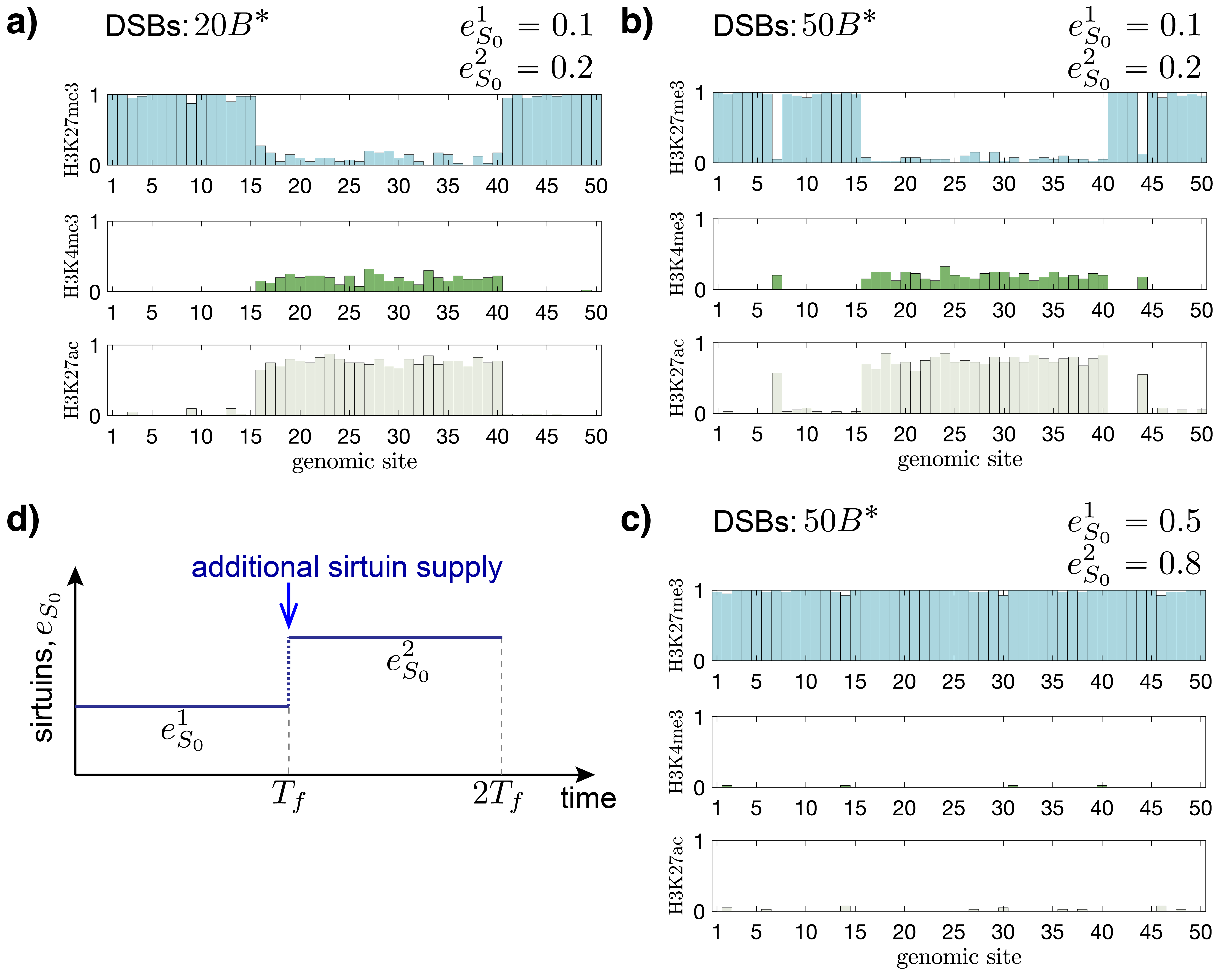}
\caption{\textbf{Stochastic simulations demonstrating the recovery of baseline epigenetic patterns under elevated DNA damage conditions through increased sirtuin levels.} \textbf{a-b)} The disrupted epigenetic states observed at low sirtuin concentration ($e_S = 0.1$) and elevated DSBs $B = 20B^*$ and $B = 50B^*$ (see Figs.~\ref{Figure3}g and i of the main text) were restored to the original rugged pattern 2 by increasing $e_S$ to 0.2 (compare with Fig.~\ref{Figure3}e). \textbf{c)} The closed chromatin state destabilised at $e_S = 0.5$ under $B = 50B^*$ (Fig.~\ref{Figure3}j) was recovered by raising the sirtuin level to $e_S = 0.8$ (compare with Fig.~\ref{Figure3}f). \textbf{d)} Schematic summary of the numerical experiments presented in \textbf{a-c}. We first simulated our stochastic model up to time $T_f = 30000$, using initial sirtuin concentrations $e_{S_0}^1$, as in Fig.~\ref{Figure3}. At time $T_f$, the sirtuin concentration was increased to $e_{S_0}^2 > e_{S_0}^1$, and the simulation was continued until time $2T_f $. The epigenetic patterns shown in \textbf{a-c} correspond to the system state at this final time point, $2T_f $. Simulation details are provided in Section~\ref{SM:sec12}; all remaining parameters are listed in Supplementary Table~\ref{SuppTable1}.}
\label{SuppFig2}
\end{center}
\end{figure}

\begin{figure}[!h]
\begin{adjustwidth}{-0.5cm}{-0.5cm}
\begin{center}
\vspace{-0.3cm}
\includegraphics[width=19cm]{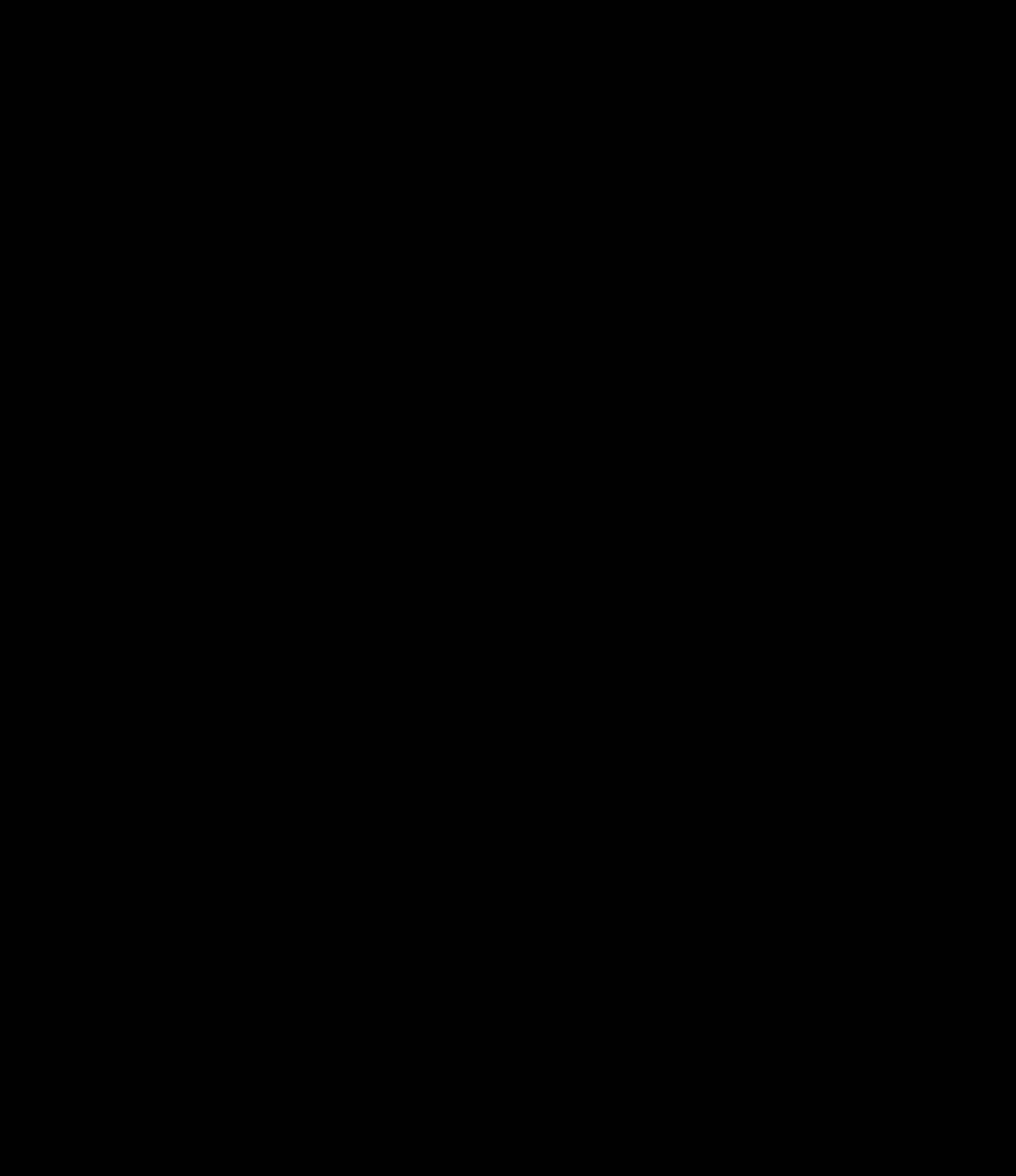}
\caption{\textbf{Geometry-dependent epigenetic stability under increasing DNA damage at high sirtuin concentrations.} \textbf{a-d)} Phase diagrams for the two-region chromatin system under high sirtuin availability ($e_S = 0.5$), displaying the number and nature of stable steady states as a function of TAD-like domain size, $\bar{R}$, and interaction strength within that domain, $w_{TAD}$, across increasing levels of DNA damage: \textbf{a)} baseline damage, $B = B^*$, \textbf{b)} 20-fold, \textbf{c)} 50-fold, and \textbf{d)} 100-fold elevations in DSBs. Diagonal interaction strength is set at $w_{diag} = 100$, with a total of $N = 50$ genomic sites. The red curve in each panel marks the condition $\bar{R} \cdot w_{TAD} = 450$, ensuring constant feedback potential across geometries. Colour scheme and hatching follow those in Figs.~\ref{Figure3}a-c (main text). The underlying mean-field framework is detailed in Section~\ref{SM:sec14}. \textbf{e-f)} Chromatin configurations used in the stochastic simulations (marked as {\protect\circled{1}} and {\protect\circled{2}} in panel \textbf{a}), both satisfying $\bar{R} \cdot w_{TAD} = 450$ but differing in topology: geometry 1 features a broad TAD region with weaker internal connectivity (\textbf{e}: $\bar{R} = 45$, $w_{TAD} = 10$), whereas geometry 2 consists of a smaller but more densely interconnected TAD domain (\textbf{f}: $\bar{R} = 5$, $w_{TAD} = 90$). \textbf{g, i, k)} Final epigenetic patterns resulting from stochastic simulations of geometry 1 under \textbf{g)} baseline, \textbf{i)} 20-fold, and \textbf{k)} 50-fold DSB conditions. Results for geometry 2 under the same damage levels are shown in \textbf{h, j, l)}. Full simulation methodology is described in Section~\ref{SM:sec12}; parameter values are listed in Supplementary Table~\ref{SuppTable1}.}
\label{SuppFig3}
\end{center}
\end{adjustwidth}
\end{figure}

\begin{figure}[!h]
\begin{adjustwidth}{-0.5cm}{-0.5cm}
\begin{center}
\vspace{-0.2cm}
\includegraphics[width=19cm]{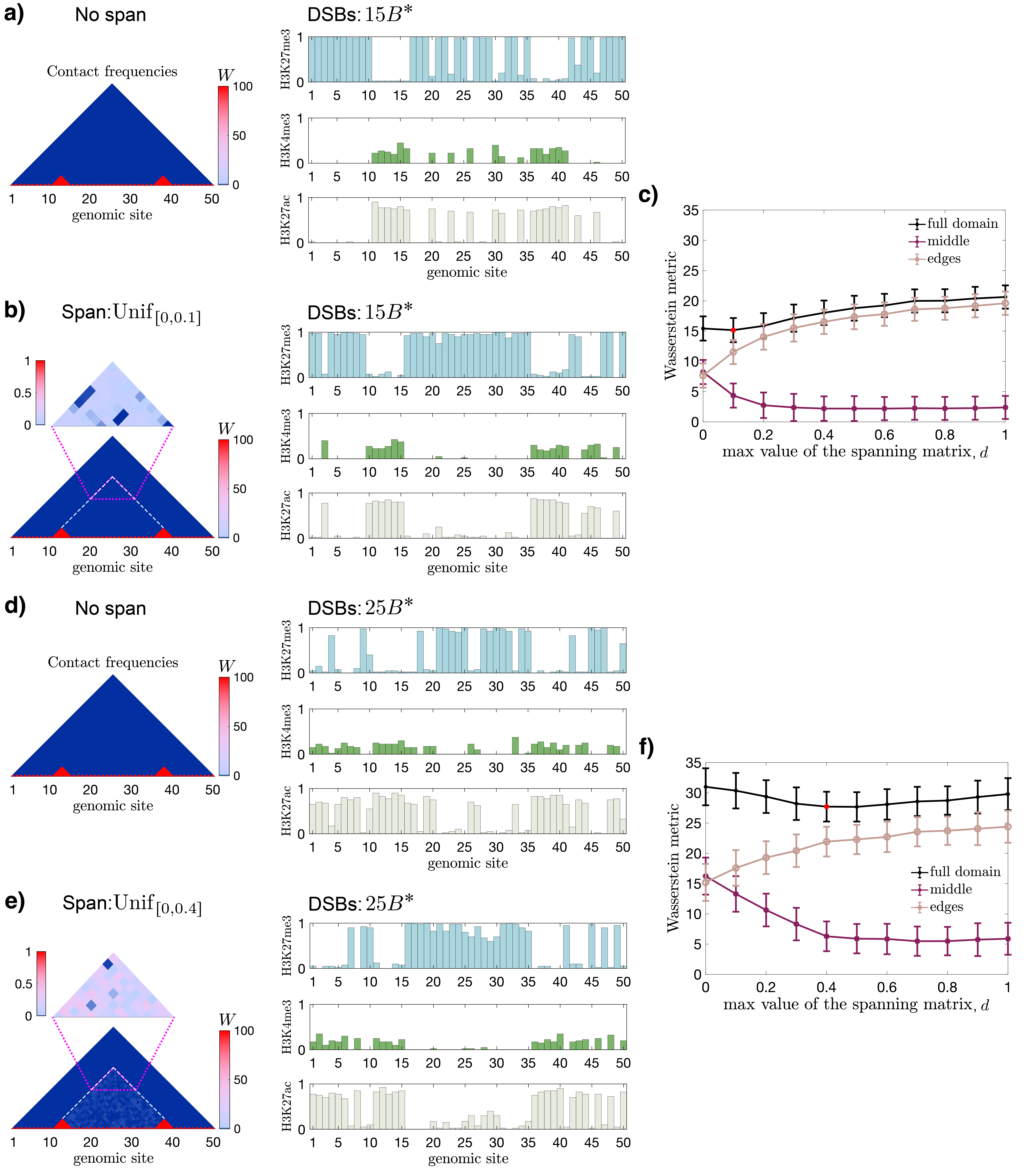}
\caption{\textbf{Stochastic simulations of the multi-site model under moderate increases in DNA damage.} \textbf{a, d)} Representative epigenetic landscapes for the chromatin geometry with two contact-rich regions separated by diagonal connections, under \textbf{a)} a 15-fold and \textbf{d)} a 25-fold increase in DSBs. These simulations show destabilised epigenetic patterns as compared to the baseline configuration under baseline DSB level, $B = B^*$ (see Fig.~\ref{Figure5}b of the main text). \textbf{b, e)} Partial recovery of the original epigenetic landscape achieved by introducing a spanning matrix of low-intensity contacts around the two hubs. Here, we show representative results for \textbf{b)} $B = 15B^*$ with $d = 0.1$, and \textbf{e)} $B = 25B^*$ with $d = 0.4$. These $d$ values correspond to the minima of the Wasserstein metric curves shown in \textbf{c, f} (full domain). \textbf{c, f)} Wasserstein distances between the reference baseline pattern and simulations with the spanning matrix at varying maximum contact values $d$, for \textbf{c)} $B = 15B^*$ and \textbf{f)} $B = 25B^*$. Metrics are computed for the full domain (sites 1-50, black), the central region (11-40, magenta), and the flanks (1-10 and 41-50, beige). Red dots mark the optimal $d$ value that minimises the divergence from the baseline. All metrics are computed using the Python Optimal Transport library \supercite{flamary2021pot}, with error bars showing standard deviation across 500 realisations.}
\label{SuppFig4}
\end{center}
\end{adjustwidth}
\end{figure}

\begin{figure}[!h]
\begin{adjustwidth}{-0.5cm}{-0.5cm}
\begin{center}
\includegraphics[width=19cm]{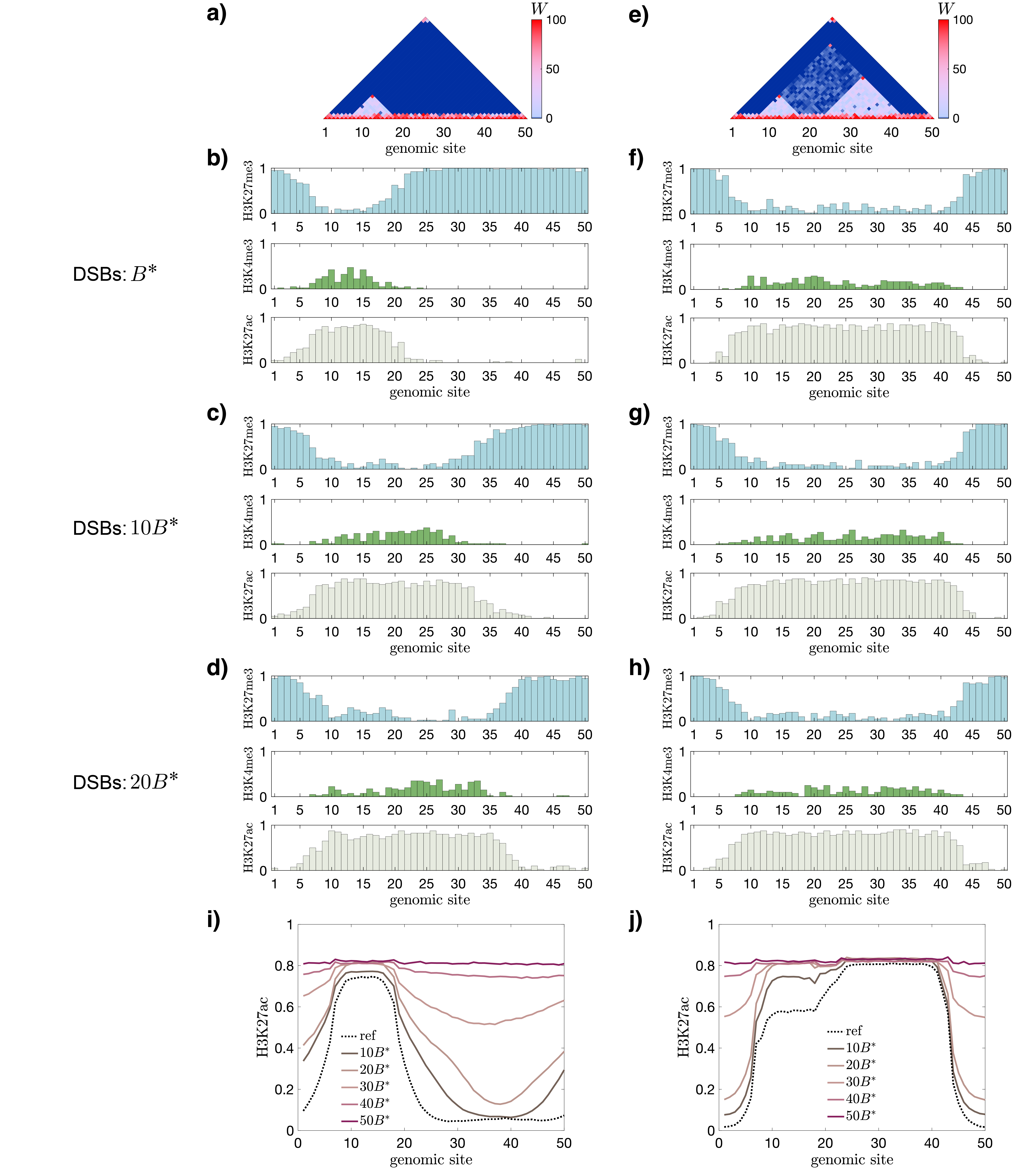}
\caption{\textbf{Stochastic simulations of epigenetic landscapes for more realistic chromatin architectures} (variation 1; differences with Fig.~\ref{Figure6} of the main text emphasised in \textcolor{blue}{blue}). \textbf{a-d)} Representative epigenetic patterns for a chromatin configuration featuring a locally coiled DNA polymer with a single small TAD domain shown in \textbf{a)}. The resulting landscapes are shown for \textbf{b)} baseline DSB levels, $B = B^*$, and for elevated damage: \textbf{c)} 10-fold and \textbf{d)} 20-fold increases. \textcolor{blue}{Contact values within the TAD region are sampled from $\unif_{[0,30]}$}, while diagonal elements are sampled from $\unif_{[60,100]}$. \textbf{e-h)} Representative epigenetic patterns for a chromatin architecture with two small TAD domains embedded within a larger region of low-intensity contacts, shown in \textbf{e)}. Corresponding epigenetic landscapes are shown for \textbf{f)} baseline DSBs, \textbf{g)} 10-fold, and \textbf{h)} 20-fold DSB increases. \textcolor{blue}{Contact values within the TAD regions are sampled from $\unif_{[0,30]}$, the surrounding embedding region from $\unif_{[0,1]}$}, and diagonal values from $\unif_{[60,100]}$. \textbf{i, j)} Averaged H3K27ac patterns for the chromatin geometries in \textbf{a} and \textbf{e}, respectively, under increasing DSB levels. Dashed black lines show the average landscape at baseline ($B = B^*$), while solid lines correspond to elevated DSBs ($10B^* \leq B \leq 50B^*$; see colour legend). Results are averaged over 500 stochastic realisations for each condition. Simulation details are provided in Section~\ref{SM:sec12}; all parameters are listed in Supplementary Table~\ref{SuppTable1}.}
\label{SuppFig5}
\end{center}
\end{adjustwidth}
\end{figure}

\begin{figure}[!h]
\begin{adjustwidth}{-0.5cm}{-0.5cm}
\begin{center}
\includegraphics[width=19cm]{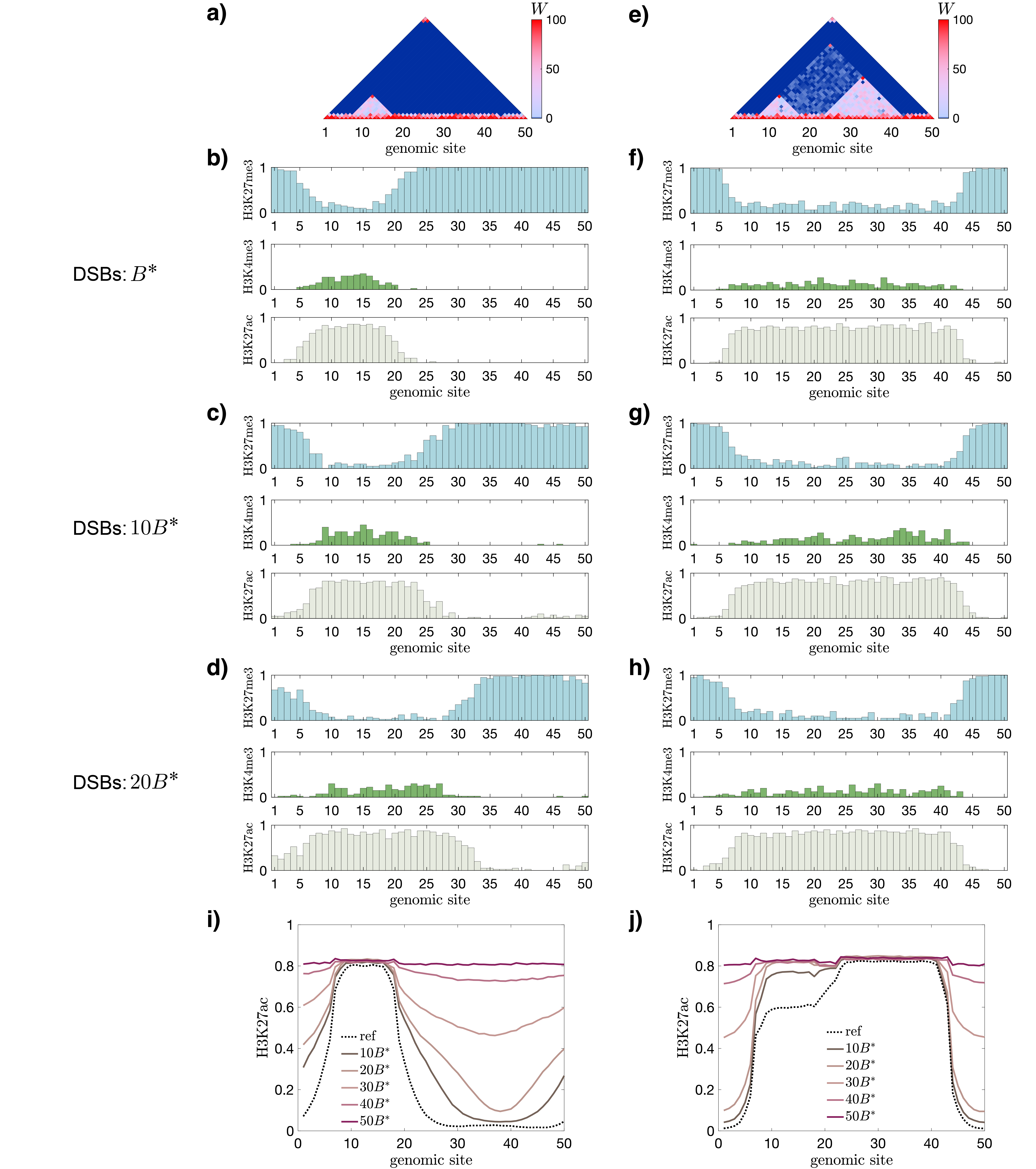}
\caption{\textbf{Stochastic simulations of epigenetic landscapes for more realistic chromatin architectures} (variation 2; differences with Fig.~\ref{Figure6} of the main text emphasised in blue). \textbf{a-d)} Representative epigenetic patterns for a chromatin configuration featuring a locally coiled DNA polymer with a single small TAD domain shown in \textbf{a)}. The resulting landscapes are shown for \textbf{b)} baseline DSB levels, $B = B^*$, and for elevated damage: \textbf{c)} 10-fold and \textbf{d)} 20-fold increases. \textcolor{blue}{Contact values within the TAD region are sampled from $\unif_{[0,50]}$}, while diagonal elements are sampled from $\unif_{[60,100]}$. \textbf{e-h)} Representative epigenetic patterns for a chromatin architecture with two small TAD domains embedded within a larger region of low-intensity contacts, shown in \textbf{e)}. Corresponding epigenetic landscapes are shown for \textbf{f)} baseline DSBs, \textbf{g)} 10-fold, and \textbf{h)} 20-fold DSB increases. \textcolor{blue}{Contact values within the TAD regions are sampled from $\unif_{[0,50]}$, the surrounding embedding region from $\unif_{[0,1]}$}, and diagonal values from $\unif_{[60,100]}$. \textbf{i, j)} Averaged H3K27ac patterns for the chromatin geometries in \textbf{a} and \textbf{e}, respectively, under increasing DSB levels. Dashed black lines show the average landscape at baseline ($B = B^*$), while solid lines correspond to elevated DSBs ($10B^* \leq B \leq 50B^*$; see colour legend). Results are averaged over 500 stochastic realisations for each condition. Simulation details are provided in Section~\ref{SM:sec12}; all parameters are listed in Supplementary Table~\ref{SuppTable1}.}
\label{SuppFig6}
\end{center}
\end{adjustwidth}
\end{figure}

\clearpage

\renewcommand{\thetable}{\arabic{table}}
\renewcommand{\tablename}{Supplementary Table}
\setcounter{table}{0}

\renewcommand{\arraystretch}{1.5}
\setlength{\doublerulesep}{2\arrayrulewidth}
%\begin{table}[!h]
%%\vspace{-0.6cm}
%\begin{tabular}{| p{1cm} | p{1.5cm} | p{12.8cm} |}
\begin{longtable}{| p{1cm} | p{1.7cm} | p{11.0cm} |}
\caption{\textbf{Baseline parameter values utilised for model analysis and simulations in this study.} The majority of these values remain consistent with those from our previous work \supercite{stepanova2025understanding} and are recalled here for reference.} \label{SuppTable1} \\
\endfirsthead
\caption{\textit{Continued.}} \\
\endhead
%\endlastfoot
\hline
\textbf{Par.} & \textbf{Value} & \textbf{Description} \\ \hline \hline
$e_{Z_0}$ & $0.75$ & Dimensionless parameter representing the relative abundance of EZH2, a methyltransferase responsible for catalysing H3K27 trimethylation. \\
$e_{U_0}$ & $0.5$ & Dimensionless parameter indicating the level of UTX, a demethylase involved in the removal of H3K27me3 marks. \\
$e_{M_0}$ & $0.1$ & Dimensionless parameter corresponding to the concentration of MLL, a methyltransferase that facilitates H3K4 trimethylation. \\
$e_{K_0}$ & $0.5$ & Dimensionless parameter quantifying the abundance of KDM, a histone lysine demethylase that removes H3K4me3 marks. \\
$e_{C_0}$ & $1.0$ & Dimensionless parameter denoting the level of CBP, an acetyltransferase that catalyses H3K27 acetylation. \\
$e_{S_0}$ & $0.1$ & Dimensionless parameter representing the presence of NuRD and sirtuins, which act as histone deacetylases. \\ \hline
$\kappa_1$ & $1.0$ & Dimensionless parameter characterising the strength of positive feedback loops promoting the deposition of H3K27me3 marks due to already existing marks of the same type (self-reinforcement). \\
$\kappa_2$ & $0.2$ & Dimensionless parameter defining the extent of inhibitory feedback exerted by H3K4me3 on H3K27me3 formation. \\
$\kappa_4$ & $1.0$ & Dimensionless parameter reflecting the reinforcing influence of H3K27ac on the establishment of H3K27me3 marks. \\
$\kappa_5$ & $1.0$ & Dimensionless parameter measuring the inhibitory feedback effect of H3K27me3 on H3K4me3 accumulation. \\ \hline
$s_{{27}_0}$ & $1.0$ & Dimensionless parameter specifying the number of H3K27 lysine residues per genomic locus (normalised to a given DNA segment length in kb). \\
$s_{{4}_0}$ & $1.0$ & Dimensionless parameter specifying the number of H3K4 lysine residues per genomic locus (normalised to a given DNA segment length in kb). \\ \hline
$\kappa_3$ & $1.0$ & Dimensionless rate constant governing the trimethylation of H3K27 by EZH2. \\
$\kappa_{12}$ & $1.0$ & Dimensionless rate constant associated with the demethylation of H3K27me3 by UTX. \\
$\kappa_6$ & $1.0$ & Dimensionless rate constant governing the trimethylation of H3K4 by MLL. \\
$\kappa_{15}$ & $1.0$ & Dimensionless rate constant for the removal of H3K4me3 marks by KDM. \\
$\kappa_9$ & $1.0$ & Dimensionless rate constant controlling the acetylation of H3K27 by CBP. \\
$\kappa_{18}$ & $1.0$ & Dimensionless rate constant for the deacetylation of H3K27ac by NuRD and sirtuins. \\ \hline
$\kappa_2 \kappa_{2_0}$ & $0.0$ & Dimensionless parameter indicating the baseline dissociation rate of an EZH2-H3K27 complex. \\
$\kappa_5 \kappa_{5_0}$ & $0.0$ & Dimensionless parameter defining the baseline dissociation rate of an MLL-H3K4 complex. \\
$\kappa_8$ & $1.0$ & Dimensionless parameter representing the baseline dissociation rate of a CBP-H3K27 complex. \\ 
$\overline{\kappa}_3$ & $\kappa_3 + \kappa_2 \kappa_{2_0}$ & Dimensionless parameter describing the combined rate of EZH2-H3K27 complex reduction due to either complex dissociation or H3K27me3 deposition. \\
$\overline{\kappa}_6$ & $\kappa_6 + \kappa_5 \kappa_{5_0}$ & Dimensionless parameter defining the rate of MLL-H3K4 complex reduction through dissociation or H3K4me3 deposition. \\
$\alpha_8$ & $\kappa_8 + \kappa_9$ & Dimensionless parameter accounting for the rate of CBP-H3K27 complex loss through dissociation or H3K27ac deposition. \\ \hline
$\alpha_{10}$ & $1.0$ & Inverse of the Michaelis constant \supercite{ingalls2013mathematical} associated with enzymatic removal of H3K27me3. \\
$\alpha_{13}$ & $1.0$ & Inverse of the Michaelis constant \supercite{ingalls2013mathematical} related to enzymatic removal of H3K4me3. \\
$\alpha_{16}$ & $1.0$ & Inverse of the Michaelis constant \supercite{ingalls2013mathematical} describing enzymatic removal of H3K27ac. \\ \hline
$\alpha_{19}$ & $1.0$ & Inverse of the dissociation constant \supercite{ingalls2013mathematical} for the complex formation between sirtuins and double-strand break sites. \\
$B$ & $0.1$ & Dimensionless parameter representing the degree of genomic instability in the system, quantified by the level of double-strand breaks. The baseline value, $B = B^* = 0.1$, corresponds to endogenous DNA damage arising from factors such as cell proliferation and environmental influences \supercite{vilenchik2003endogenous}. \\
\hline
$\alpha_{I_0}$ & $0.1$ & Dimensionless parameter specifying the baseline formation rate of EZH2-H3K27 complexes. \\
$\alpha_{A_0}$ & $1.0$ & Dimensionless parameter indicating the baseline formation rate of MLL-H3K4 complexes. \\
$\alpha_{Q_0}$ & $0.1$ & Dimensionless parameter corresponding to the baseline formation rate of CBP-H3K27 complexes. \\ \hline
%\end{tabular}
%\vspace{0.1cm}
%\caption{caption} \label{TableBaselineParams}
\end{longtable}

\makeatletter\@input{xxsupp.tex}\makeatother